\documentclass[12pt]{article}
\usepackage{redo-cl-sit}
\begin{document}
\title{Shannon Information Theory \\
Without Shedding Tears\\
Over Delta \& Epsilon Proofs\\
or Typical Sequences}

\author{Robert R. Tucci\\
        P.O. Box 226\\
        Bedford,  MA   01730\\
        tucci@ar-tiste.com}

\date{\today}
\maketitle
\vskip2cm
\section*{Abstract}
This paper begins with a
discussion of integration
over probability types (p-types).
After doing that, the paper re-visits
3 mainstay problems of classical (non-quantum)
Shannon Information Theory (SIT):
source coding without distortion,
channel coding,
and source coding with distortion.
The paper proves
well-known, conventional results for
each of these 3 problems. However,
the proofs given for these results
are not conventional. They
are based on complex integration techniques
(approximations obtained by applying
the method of steepest descent to
p-type integrals)
instead of the usual delta \& epsilon
and typical sequences
arguments.
Another
unconventional feature
of this paper
is that we make ample
use of classical Bayesian networks (CB nets).
This paper showcases
some of the benefits of using
CB nets to do classical SIT.

\newpage

\section{Introduction}

For a good textbook on classical (non-quantum)
Shannon
Information Theory (SIT), see, for example,
Ref.\cite{CovTh}
by Cover and Thomas.
Henceforth we will refer to
it
as C\&T.
For a good textbook on classical (non-quantum)
Bayesian Networks, see, for example,
Ref.\cite{KF}
by Koller and Friedman.

This paper begins with a
discussion of integration
over probability types (p-types).
After doing that, the paper re-visits
3 mainstay problems of classical
SIT:

\begin{itemize}
\item source coding (lossy compression)
without distortion
\item channel coding
\item source coding with distortion
\end{itemize}
The paper proves
well-known, conventional results for
each of these 3 problems. However,
the proofs given for these results
are not conventional. They
are based on complex integration techniques
(approximations obtained by applying
the method of steepest descent to
p-type integrals)
instead of the usual delta \& epsilon
and typical sequences
arguments.

Another
unconventional feature
of this paper
is that we make ample
use of classical Bayesian networks (CB nets).
This paper showcases
some of the benefits of using
CB nets to do classical SIT.

P-types were introduce into SIT
by Csisz\'{a}r and K\"{o}rner (see Ref.\cite{CK}).
P-type integration is
a natural, almost obvious
consequence of the theory of
p-types,
although it is not spelled out
explicitly in the book
by Csisz\'{a}r and K\"{o}rner.
In fact, all workers
whose work I am familiar
with, including Csisz\'{a}r and K\"{o}rner,
use p-types frequently, but
they do not use p-type integration.
Instead, they use delta \& epsilon
and typical sequences
arguments to bound some finite sums
which are discrete approximations
of p-type integrals.

The conventional delta \& epsilon arguments
are more rigorous than
the p-type integration arguments
presented here.
Although less rigorous
than traditional arguments,
p-type integration arguments
have the virtue that they are
easier to understand and follow,
especially
by people who are not well versed
in rigorous analysis.
Such is the case with many
physicists and engineers.
A similar problem occurs
when teaching Calculus.
One can teach
Calculus with the full panoply of
delta \& epsilon arguments
from a textbook such as the
legendary one by W. Rudin (Ref.\cite{Rudin}).
Or one can teach Calculus
at the level and scope of
 a college freshman  course
for engineers.
Each approach appeals to
a different audience and fulfils different
needs.

Most of our results are not
exact. They are
leading order terms
in asymptotic expansions
for large $n$,
where $n$ is the number of letters
in a codeword. These
approximations become increasingly
more accurate as $n\rarrow \infty$.

This paper is almost self contained,
although a few times we
assume
certain inequalities
and send the reader to C\&T for a proof
of them.

\section{Preliminaries and Notation}

In this section, we
will describe
some basic notation
used throughout this paper.

As usual, $\ZZ,\RR, \CC$
will denote the integers, real numbers,
and complex numbers, respectively.
We will sometimes
add superscripts to these
symbols to indicate subsets of
these sets. For instance,
we'll use $\RR^{\geq 0}$
to denote the set of non-negative reals.
For $a,b\in\ZZ$ such that $a\leq b$,
let
 $Z_{a,b}=\{a,a+1,a+2,\ldots, b\}$.

 Let
$\delta^x_{y}=\delta(x,y)$
denote the Kronecker delta function:
it equals 1 if $x=y$ and 0 if $x\neq y$.
Let $\theta(S)$ denote the truth function:
it equals 1 if statement
$S$ is true and 0 otherwise.
For example, $\delta_x^y = \theta(x=y)$.
Another example is the step function $\theta(x>0)$:
it equals 1 if $x>0$ and is zero otherwise.

For any matrix $M\in \CC^{p\times q}$,
$M^*$ will denote its complex conjugate,
$M^T$ its transpose, and $M^\dagger = M^{*T}$
its Hermitian conjugate.

Random variables will be denoted
by underlined letters; e.g.,
$\rva$.
The (finite) set of values (states) that
$\rva$ can assume will be denoted
by $S_\rva$. Let $N_\rva=|S_\rva|$.
The probability that
$\rva=a$ will be denoted by $P(\rva=a)$
or $P_\rva(a)$, or simply by $P(a)$
if the latter will not lead to confusion
in the context it is being used.
We will use $pd(S_\rva)$ to denote
the set of all probability distributions
with domain $S_\rva$. For joint random variables
$(\rva,\rvb)$, let
$S_{\rva,\rvb} =
S_\rva\times S_\rvb=
\{(a,b):a\in S_\rva, b\in S_\rvb\}$.

Sometimes, when
two random variables $\rva\av{1}$ and
$\rva\av{2}$ satisfy
$S_{\rva\av{1}}=S_{\rva\av{2}}$,
we will omit the indices $\av{1}$
and $\av{2}$ and refer to both
random variables as $\rva$.
We shall do this sometimes
even if the
random variables $\rva\av{1}$
 and $\rva\av{2}$ are not identically distributed!
This notation, {\it if used with caution},
does not lead to confusion and
does avoid a lot of index clutter.

Suppose
$\{P_{\rvx,\rvy}(x,y)\}_{\forall x,y}
\in pd(S_{\rvx,\rvy})$.
We will often use
the expectation operators
$E_x = \sum_x P(x)$,
 $E_{x,y}=\sum_{x,y} P(x,y)$,
 and
$E_{y|x}=\sum_y P(y|x)$.
Note that $E_{x,y} = E_x E_{y|x}$.
Let
\beq
P(x:y) = \frac{P(x,y)}{P(x)P(y)}
\;.
\eeq
Note that
$E_{x} P(x:y) = E_{y} P(x:y)=1$.

Suppose
$n$ is any positive integer.
Let $\rvx^n = (\rvx_1, \rvx_2, \ldots, \rvx_n)$
be the random variable that takes one values
$x^n = (x_1, x_2, \ldots, x_n)\in S_\rvx^n$.

The {\bf rate of $\rvx$} is defined as
$R_\rvx = \frac{\ln N_\rvx}{n}$.

$\rvx^n$ is said to be i.i.d.
(independent, identically
distributed)
if $S_{\rvx_j}=S_\rvx$ for all $j\in Z_{1,n}$
and there is a $P_\rvx\in pd(S_\rvx)$
such that
$P_{\rvx^n}(x^n)=
\prod_{j=1}^n\{
P_\rvx(x_j)\}$.
When $\rvx^n$ is i.i.d.,
we will sometimes use
$P_{\rvx}(x^n)$ to denote
the more correct expression
$P_{\rvx^n}(x^n)$
and say that
$P_\rvx(x^n)$
is an i.i.d. source.

Suppose $\{P(y^n|x^n)\}_{\forall y^n}\in pd(S^n_\rvy)$
for all $x^n\in S_\rvx^n$.
$P(y^n|x^n)$
is said to be a discrete memoryless channel (DMC)
if
$P(y^n|x^n) =\prod_{j=1}^{n}P(y_j|x_j)$.

We will use the following
measures of various
types of information (entropy):

\begin{itemize}
\item
The (plain) entropy of the
random variable $\rvx$ is defined
in the classical case by

\beq
H(\rvx) =
E_x \ln \frac{1}{P(x)}
\;,
\eeq
which we also call
$H_{P_\rvx}(\rvx)$,
$H\{P(x)\}_{\forall x}$,
and
$H(P_\rvx)$.
This quantity measures the
spread of $P_\rvx$.

One can also consider
plain entropy for
a joint random variable
$\rvx=(\rvx_1,\rvx_2)$.
For $P_{\rvx_1,\rvx_2}\in pd(S_{\rvx_1,\rvx_2})$
with marginal probability distributions $P_{\rvx_1}$
and $P_{\rvx_2}$,
one defines a joint entropy $H(\rvx_1,\rvx_2)=H(\rvx)$
and partial entropies
$H(\rvx_1)$ and $H(\rvx_2)$.

\item
The conditional entropy of $\rvy$ given $\rvx$
is defined
in the classical case by

\beqa
H(\rvy|\rvx) &=&
E_{x,y} \ln \frac{1}{P(y|x)}
\\
&=&
H(\rvy,\rvx)-H(\rvx)
\;,
\eeqa
which we also call
$H_{P_{\rvx, \rvy}}(\rvy|\rvx)$.
This quantity measures  the conditional
 spread
of $\rvy$ given $\rvx$.

\item The Mutual Information (MI)
of $\rvx$ and $\rvy$
is defined
in the classical case by

\beqa
H(\rvy:\rvx) &=&
E_{x,y} \ln
P(x:y)= E_x E_y P(x:y) \ln P(x:y)
\\
&=&
H(\rvx) + H(\rvy) - H(\rvy,\rvx)
\;,
\eeqa
which we also call
$H_{P_{\rvx,\rvy}}(\rvy:\rvx)$.
This quantity measures the correlation
between $\rvx$ and $\rvy$.
\item The Conditional Mutual Information (CMI,
which can be read as ``see me")
of $\rvx$ and $\rvy$
given $\rv{\lam}$
is defined
in the classical case by:

\beqa
H(\rvy:\rvx|\rv{\lam})
&=&
E_{x,y,\lam} \ln
\frac{P(x,y|\lam)}{P(x|\lam)P(y|\lam)}
\\
&=&
E_{x,y,\lam} \ln
\frac{P(x,y,\lam)P(\lam)}{P(x,\lam)P(y,\lam)}
\\
&=&
H(\rvx|\rv{\lam}) + H(\rvy|\rv{\lam})
- H(\rvy,\rvx|\rv{\lam})
\;,
\eeqa
which we also call
$H_{P_{\rvx,\rvy,\rv{\lam}}}(\rvy:\rvx|\rv{\lam})$.
This
quantity measures the conditional correlation
of $\rvx$ and $\rvy$ given $\rv{\lam}$.

\item The relative
information
of $P\in pd(S_\rvx)$
divided by $Q\in pd(S_\rvx)$
is defined by

\beq
D\{P(x)//Q(x)\}_{\forall x} =
\sum_x P(x)\ln\frac{P(x)}{Q(x)}
\;,
\eeq
which we also call
$D(P_{\rvx}//Q_\rvx)$.

\end{itemize}

Note that we
define entropies
using natural logs. Our
strategy is to
use natural log entropies
for all intermediate analytical
calculations, and to
convert
to base-2 logs
at the end of those
calculations if
a base-2 log numerical answer
is desired. Such a conversion is
of course trivial
using $\log_2 x = \frac{\ln x}{\ln 2}$ and
$\ln 2 = 0.6931$

We  will use
the following well-known integral representation
of the Dirac delta function:

\beq
\delta(x) =
\int_{-\infty}^{+\infty}
\frac{dk}{2\pi}\;\;  e^{ikx}
\;.
\eeq
We will also use the following
 integral
representation
of the step function:

\beq
\theta(x>0) =
\int_{-\infty}^{+\infty}
\frac{dk}{2\pi i}\;\;
\frac{e^{ikx}}{(k-i\eps)}
\;,\label{eq-theta-comp-int}
\eeq
for some $\eps>0$.
Eq.(\ref{eq-theta-comp-int})
follows
because the integrand
has a simple pole at $k=i\eps$.
Let $k = k_r + i k_i$.
If $x>0$,
the integrand goes to zero
in the upper
half of the $(k_r,k_i)$ plane
and it goes to infinity
in the lower half plane, so we
are forced to close the contour
of integration in the upper half plane,
which means the pole lies inside the contour.
When $x<0$,
we are forced to close the contour in the lower
half plane and thus the pole lies outside
the contour.

Suppose $\call(v)$
is a real valued
function
that depends in a continuous
manner on $N$
real variables $v=\{v_j\}_{j=1}^N$.
The following variational
operator can be applied to $\call(v)$:

\beq
\delta= \sum_j \delta v_j
\pder{\;\;}{v_j}
\;.
\eeq
The $N$-dimensional Taylor
expansion of $\call(v)$
about the point $v=0$ can be expressed as

\beq
f(v) = f(0) +[\delta f(v)]_{v=0} +\frac{1}{2!}
[\delta^2 f(v)]_{v=0} + +\frac{1}{3!}
[\delta^3 f(v)]_{v=0} +\ldots
\;.
\eeq

We will often use the following
Taylor expansions:

\beq
x^\eps = e^{\eps\ln x}= 1 + \eps \ln x +
\frac{1}{2}(\eps \ln x)^2 + \ldots
\;,
\label{eq-taylor-one}
\eeq
and

\beq
\ln(1+x)= x - \frac{x^2}{2} + \ldots
\; (\mbox{ converges if }|x|<1)
\;.
\label{eq-taylor-two}
\eeq

\section{Integration Over P-types}
In this section, we will define
integration over probability types (p-types).
The set of p-types for a
given $n$
fills all of
$pd(S_\rvx)$
in an increasingly
finer way as $n\rarrow \infty$. Thus, once
the density
of p-types at
each
point of $pd(S_\rvx)$ is known,
we can integrate that density
 over
a particular region $R\subset pd(S_\rvx)$
to get the
number of p-types
within $R$.
We will define
integration over p-types
that depend on a single
variable (univariate p-types),
or multiple variables (multivariate p-types).
We will also define integration over
conditional p-types.
Finally, we will define Dirac
delta functions for integration
over p-types.

\subsection{Integration Over Univariate P-type}

For any $x^n\in S^n_\rvx$,
denote the number of occurrences
of $x\in S_\rvx$ within $x^n$
by $N(x|x^n)$. Hence

\beq
N(x|x^n) = \sum_{j=1}^n
\theta(x_j=x)
\;.
\eeq
One can now say that two elements $x^n$ and
$x'^n$
of $S_\rvx^n$ are
equivalent if, for all $x\in S_\rvx$,
$x^n$ and $x'^n$ both have the same
number of occurrences
of $x$. This equivalence relation
partitions $S^n_\rvx$
into equivalence classes given
by, for any $x^n\in S_\rvx^n$,

\beq
[x^n] = \{x'^n\in S_\rvx^n:
N(x|x^n)= N(x|x'^n) \forall x\in S_\rvx
\}
\;.
\eeq
For each class
$[x^n]$
and $x\in S_\rvx$, we
can define

\beq
\ptype{x^n}(x) =
\frac{N(x|x^n)}{n}
\;.
\eeq
Clearly,
$\{\ptype{x^n}(x)\}_{\forall x}\in pd(S_\rvx)$.
We will refer to
this probability distribution as a p-type.

Note that if $Q(x^n)$
is an i.i.d. source,

\beq
Q(x^n) = \prod_{j=1}^n Q(x_j)
\;,
\eeq
so

\beq
Q(x^n) =
\prod_{x\in S_\rvx}
\left\{
Q(x)^{N(x|x^n)}\right\}=
e^{
n\sum_x
\ptype{x^n}(x)
\ln Q(x)
}
\;.
\eeq

Define the following integration operator:

\beq
\int \cald \ptype{x^n}
=
\prod_x
\left\{
\int_0^1 d\ptype{x^n}(x)
\right\}
\delta\left(\sum_x \ptype{x^n}(x) -1\right)
\;.
\label{eq-p-type-int-measure}
\eeq

We will denote the number of elements in
a class $[x^n]$ by

\beq
d_{[x^n]} = |[x^n]|
\;.
\eeq

\begin{claim}

\beq
\sum_{x^n} =
\sum_{[x^n]} d_{[x^n]}
\;.
\eeq
\end{claim}
\proof
The classes $[x^n]$
are non-overlapping
and they cover all of $S^n_\rvx$.
\qed
\begin{claim}
For any $x^n\in S^n_\rvx$,
\beq
d_{[x^n]}=
 (d_{[x^n]})_{H=0}\;\;e^{nH(\ptype{x^n})}
\;,
\eeq
where

\beq
(d_{[x^n]})_{H=0}=
\frac{1}
{(2\pi n)^{\frac{N_\rvx-1}{2}}
\sqrt{
\prod_x \ptype{x^n}(x)}
}
\;.
\eeq
\end{claim}
\proof
Let
\beq
S_\rvx = \{x(j): j\in Z_{1,N_\rvx}\}
\;
\eeq
and

\beq
 r_j = N(x(j)|x^n)
\;
\eeq
for all $j\in Z_{1,N_\rvx}$.
Note that $\sum_{j=1}^{N_\rvx} r_j = n$.
Recall Stirling's formula:

\beq
n! \approx  \sqrt{2\pi n} \;\;n^n e^{-n}
\;
\eeq
for $n>>1$.
Combinatorics gives
a value for $|[x^n]|$ in terms of
factorials. If we approximate
those factorials using Stirling's
formula, we get

\beqa
|[x^n]| &=&
\frac{n!}
{\prod_{j=1}^{N_\rvx} \{r_j!\}}
\\
&=&
\frac{1}
{(2\pi)^{\frac{N_\rvx-1}{2}}}
\left(
\frac{n}{r_1 r_2 \ldots r_{N_\rvx}}
\right)^{\frac{1}{2}}
e^{-n + n\ln n
-\sum_j\{-r_j + r_j\ln r_j\}
}
\\
&=&
\frac{\exp(-n\sum_j \frac{r_j}{n}
\ln \frac{r_j}{n})}
{(2\pi n)^{\frac{N_\rvx-1}{2}}
\sqrt{\prod_j\left\{
\frac{r_j}{n}
\right\}}
}
\;.
\eeqa
\qed

\begin{claim}
\beq
\sum_{[x^n]} = \int \cald \ptype{x^n}n^{N_\rvx-1}
\;.
\label{eq-sum-xn}
\eeq
\end{claim}
\proof
For any i.i.d. source $Q(x^n)$, we have that

\beqa
1 &=&
\sum_{x^n}Q(x^n)
\\
&=&
\sum_{[x^n]} d_{[x^n]}
e^{n\sum_x \ptype{x^n}(x) \ln Q(x)}
\\
&=&
\int
\frac{\cald \ptype{x^n}}
{\Delta V}
\frac{
e^{\call_0}
}
{(2\pi n)^{
\frac{N_\rvx-1}{2}
\sqrt{\prod_x \ptype{x^n}(x)}}
},
\;
\label{eq-one-is-int}
\eeqa
where
$\Delta V$ is yet to be determined and

\beq
\call_0 = n
\sum_x \ptype{x^n}(x) \ln
\frac{Q(x)}{\ptype{x^n}(x)}
\;.
\eeq
We add to $\call_0$ a Lagrange multiplier
term that constrains the components
of the vector
$\{\ptype{x^n}(x)\}_{\forall x}$
so that they sum to one:

\beq
\call = \call_\lam =
\call_0 + n\lam \left(
\sum_x \ptype{x^n}(x) -1
\right)
\;
\eeq
for any $\lam\in \RR$.
Our goal is to approximate
the integral Eq.(\ref{eq-one-is-int})
using
the method of steepest descent.
We just want to get the leading order term
in an asymptotic expansion
of the integral for large $n$.
To get this leading order term,
it is sufficient to approximate
$\call$
to second order
in $\delta \ptype{x^n}(x)$, about
the point (or points)
that have a vanishing first
variation $\delta \call$.
Thus, approximate

\beq
\call \approx
\tcall + \delta \tcall
+
\frac{1}{2}
\delta^2 \tcall
\;,
\eeq
where
quantities with a tilde over
them are evaluated at a tilde (saddle)
point that
satisfies

\beq
\delta \tcall=0
\;.
\eeq
It's easy to check that

\beq
\delta \call=
n \sum_x \delta \ptype{x^n}(x)\ln
\left(
\frac{Q(x)e^{-1+\lam}}{\ptype{x^n}(x)}
\right)
\;,
\eeq
and

\beq
\delta^2\call=
-n\sum_x
\frac{[\delta \ptype{x^n}(x)]^2}{\ptype{x^n}(x)}
\;.
\eeq
Next, for each $x$,
we set to zero
the coefficient
of
$\delta \ptype{x^n}(x)$
in $\delta \call$.
After doing that, we
enforce the constraint
that $\sum_x \ptype{x^n}(x)=1$.
This leads us to conclude that

\beq
\ptiltype{x^n}(x) = Q(x)
\;.
\eeq
Using this value of
$\ptiltype{x^n}(x)$,
we get

\beq
\tcall = 0
\;
\eeq
and

\beq
\delta^2\tcall=
-n\sum_x
\frac{[\delta \ptype{x^n}(x)]^2}{Q(x)}
\;.
\eeq
From Eq.(\ref{eq-one-is-int}),
we get

\beq
1=\frac{1}{\Delta V (2\pi n)^{
\frac{N_\rvx-1}{2}
\sqrt{\prod_x Q(x)}}
}\Gamma
\;,
\eeq
where

\beq
\Gamma =\int \cald \ptype{x^n}
e^{
-n\sum_x
\frac{\left[\delta \ptype{x^n}(x)\right]^2}{2Q(x)}
}
=
\sqrt{
\frac{\pi^{N_\rvx -1}}
{
\prod_x \left\{
\frac{n}{2Q(x)}
\right\}
\frac{2}{n}
}
}
\;.
\eeq
The final integral was performed
using Eq.(\ref{eq-p-type-int-over-gauss}).
This implies $1/\Delta V= n^{N_\rvx-1}$.
\qed

Note that Eqs.(\ref{eq-sum-xn})
and (\ref{eq-caldpx-one}) imply that

\beq
\sum_{[x^n]}
1=
\frac{n^{N_\rvx-1}}{(N_\rvx-1)!}
\;
\eeq
so the number
of p-types
with a given $n$
in $pd(S_\rvx)$
varies polynomial with $n$.

\subsection{Integration Over Multivariate P-types}

There exists
a very natural 1-1 onto map from
$S_\rvx^n\times S_\rvy^n$
to $(S_\rvx\times S_\rvy)^n$,
namely the one that
identifies $(x_j)_{\forall j}(y_j)_{\forall j}$
with $\left[\begin{array}{c}x_j\\y_j
\end{array}\right]_{\forall j}$. Thus,
the definitions
and claims given in the previous
section for $N(x|x^n)$, $[x^n]$,
$\ptype{x^n}(x)$ and
$\int \cald \ptype{x^n}$
generalize very naturally to
give analogous definitions and claims
for
$N(x,y|x^n, y^n)$, $[x^n, y^n]$,
$\ptype{x^n,y^n}(x,y)$ and
$\int \cald \ptype{x^n,y^n}$.
For example,

\beq
N(x,y|x^n,y^n)=N(
\left(\begin{array}{c}x\\y\end{array}\right)
|
\left(\begin{array}{c}x^n\\y^n\end{array}\right))
=
\sum_j\theta(
\left(\begin{array}{c}x\\y\end{array}\right)
=
\left(\begin{array}{c}x_j\\y_j\end{array}\right)
)
\;.
\eeq

We will sometimes use $[\;]$
as an abbreviation for a class. For example,
we might abbreviate
$\ptype{a^n, b^n, c^n}(a,b,c)$ by
$\ptype{\;}(a,b,c)$.

Note that when $y^n=x^n$ in $\ptype{x^n,y^n}$,

\beq
\ptype{x^n,x^n}(x,y) =
\delta_{y}^{x}\; \ptype{x^n}(x)
\;.
\eeq

Note also that we can express $\delta_{x^n}^{y^n}$
as follows

\beqa
e^{n\sum_{x,y} \ptype{x^n,y^n}(x,y)
\ln \delta_y^x}
&=&
\left\{
\begin{array}{l}
\mbox{0, if } \exists(x,y)
\mbox{ such that }\delta_x^y= 0
\mbox{ and } \ptype{x^n,y^n}(x,y)\neq 0
\\
1, \mbox{ otherwise}
\end{array}
\right.
\\
&=&
\theta(\forall(x,y): y\neq x\Rightarrow
\ptype{x^n,y^n}(x,y)= 0)
\\
&=&
\delta_{x^n}^{y^n}
\;.
\eeqa
\subsection{Integration Over Conditional P-types}
 For any $x^n\in S^n_\rvx$ and
$y^n\in S^n_\rvy$, define conditional classes by

\beq
[y^n|x^n] = \left\{(x'^n,y'^n)\in
S_\rvx^n\times S_\rvy^n:
\frac{N(x,y|x^n,y^n)}{\sum_y N(x,y|x^n,y^n)}
=
\frac{N(x,y|x'^n,y'^n)}{\sum_y N(x,y|x'^n,y'^n)}
\forall (x,y)\in
S_\rvx\times S_\rvy
\right\}
\;
\eeq
and conditional probability types by

\beq
\ptype{y^n|x^n}(y|x) =
\frac{N(x,y|x^n,y^n)}{\sum_y N(x,y|x^n, y^n)}
=
\frac{\ptype{x^n,y^n}(x,y)}{\ptype{x^n,y^n}(x)}
\;
\eeq
for all $x\in S_\rvx$ and $y\in S_\rvy$.

We will sometimes use $[\;]$
as an abbreviation for a
conditional class. For example,
we might abbreviate
$\ptype{a^n, b^n| c^n, d^n}(a,b|c, d)$ by
$\ptype{\;}(a,b|c, d)$.

Define the following  integration operator:

\beq
\int \cald \ptype{y^n|x^n}
=\prod_{x,y}
\left\{
\int_0^1 d\ptype{y^n|x^n}(y|x)
\right\}
\prod_x
\left\{
\delta\left(\sum_y \ptype{y^n|x^n}(y|x) -1
\right)
\right\}
\;.
\label{eq-cond-p-type-int-measure}
\eeq

We will denote the number of elements
in conditional class $[y^n|x^n]$ by

\beq
d_{[y^n|x^n]}=
|[y^n|x^n]|
\;.
\eeq

\begin{claim}
\label{claim-sum-xn-yn}
\beq
\sum_{x^n,y^n}=
\sum_{[x^n]}d_{[x^n]}
\sum_{[y^n|x^n]}d_{[y^n|x^n]}
\;.
\eeq
\end{claim}
\proof
For any DMC $Q(y^n|x^n)$, we must have

\beq
1=\sum_{[y^n|x^n]}d_{[y^n|x^n]}Q(y^n|x^n)
\;.
\eeq
If $Q(x^n)$ is an i.i.d source
and $Q(x^n,y^n)=Q(y^n|x^n)Q(x^n)$, then
the last equation implies

\beqa
1&=&\sum_{[x^n]}d_{[x^n]}Q(x^n)
\sum_{[y^n|x^n]}d_{[y^n|x^n]}Q(y^n|x^n)
\\
&=&
\sum_{[x^n]}d_{[x^n]}
\sum_{[y^n|x^n]}d_{[y^n|x^n]}Q(x^n,y^n)
\;.
\eeqa
But also

\beq
1=\sum_{x^n,y^n}Q(x^n,y^n)
\;.
\eeq
Since $Q(x^n,y^n)$
is an arbitrary i.i.d. source,
the claim follows.
\qed

\begin{claim}\label{claim-d-yn-cond-xn}
\beq
d_{[y^n|x^n]}=
\frac{d_{[x^n,y^n]}}{d_{[x^n]}}
\;.
\eeq
\end{claim}
\proof
Combinatorics?
\qed

\begin{claim}\label{claim-sums-chain-rule}
\beq
\sum_{[x^n]}
\sum_{[y^n|x^n]}
=
\sum_{[x^n,y^n]}
\;.
\label{eq-sums-chain-rule}
\eeq
\end{claim}
\proof
This follows from Claims \ref{claim-sum-xn-yn}
and \ref{claim-d-yn-cond-xn}
and the fact that $\sum_{x^n,y^n}=\sum_{[x^n,y^n]}
d_{[x^n,y^n]}$.
\qed

Alternatively, one could prove
Claim \ref{claim-sums-chain-rule}
by combinatorics and then
prove Claim \ref{claim-d-yn-cond-xn}
from
Claims \ref{claim-sum-xn-yn}
and \ref{claim-sums-chain-rule}.

\begin{claim}\label{claim-split-int-ptype}
\beq
\int \cald \ptype{x^n}
\int \cald \ptype{y^n|x^n}
\left[
\prod_x
\left\{
\ptype{x^n}(x)
\right\}
\right]^{N_\rvy- 1}
=
\int \cald \ptype{x^n,y^n}
\;\label{eq-split-int-ptype}
\eeq
\end{claim}
\proof
Let LHS and RHS denote
the left hand side and right hand side
of Eq.(\ref{eq-split-int-ptype}).

Recall that Dirac delta functions obey
$\delta(ax) = \frac{1}{|a|}\delta(x)$.
This proof hinges on that simple identity.

Define
\beq
\Omega_1 =
\prod_x
\left\{
\int_0^1 d\ptype{x^n}(x)
\right\}
\delta\left(\sum_x \ptype{x^n}(x) -1\right)
\;
\eeq
and

\beq
\Omega_2 =
\prod_{x,y}
\left\{
\int_0^1 d\ptype{y^n|x^n}(y|x)
\right\}
\prod_x
\left\{
\delta\left(\sum_y \ptype{y^n|x^n}(y|x) -1
\right)
\right\}
\left[
\prod_x
\left\{
\ptype{x^n}(x)
\right\}
\right]^{N_\rvy- 1}
\;.
\eeq
Then

\beqa
LHS &=& \Omega_1\Omega_2
\\
&=&\Omega_1
\prod_{x,y}
\left\{
\int_0^1 d\ptype{x^n,y^n}(x,y)
\right\}
\prod_x
\left\{
\delta\left(\sum_{y} \ptype{x^n,y^n}(x,y)
-\ptype{x^n}(x)\right)
\right\}
\\
&=&
\prod_{x,y}
\left\{
\int_0^1 d\ptype{x^n,y^n}(x,y)
\right\}
\delta\left(\sum_{x,y} \ptype{x^n,y^n}(x,y)
-1\right)
\;
\\
&=&
RHS
\;
\eeqa

This works because
LHS has $n_i=N_\rvx + N_\rvx N_\rvy$
integrals and $n_\delta=N_\rvx +1$ delta functions,
for a total of $n_i - n_\delta =
N_\rvx N_\rvy-1$ degrees of freedom.
RHS has $N_\rvx N_\rvy$ integrals and
one delta function for the
{\it same  total} of
$N_\rvx N_\rvy-1$ degrees of freedom.
\qed

\begin{claim}
\beqa
\sum_{[y^n|x^n]} &=&
\int \cald \ptype{y^n|x^n}
\frac{n^{N_\rvy N_\rvx}}
{n^{N_\rvx}}
\left[
\prod_x
\left\{
\ptype{x^n}(x)
\right\}
\right]^{N_\rvy- 1}
\\
&=&
\int \cald \ptype{y^n|x^n}
(n \ptype{x^n}^{g.m.})^{N_\rvx N_\rvy - N_\rvx}
\;,
\eeqa
where

\beq
\ptype{x^n}^{g.m.}=
\left[
\prod_x
\left\{
\ptype{x^n}(x)
\right\}
\right]^{\frac{1}{N_\rvx}}
\;
\eeq
is the geometric mean of $\ptype{x^n}$.
\end{claim}
\proof
Substitute

\beq
\int \cald \ptype{x^n}
=\frac{1}{n^{N_\rvx-1}}\sum_{[x^n]}
\;,
\eeq
and

\beq
\int \cald \ptype{x^n,y^n}
=\frac{1}{n^{N_\rvx N_\rvy-1}}
\sum_{[x^n,y^n]}
\;
\eeq
into Eq.(\ref{eq-split-int-ptype})
and then compare the result with
Eq.(\ref{eq-sums-chain-rule}).
\qed

\subsection{Dirac Delta Functions For P-type Integration}

One occasionally finds it useful to use
Dirac delta functions for p-type integration.
Suppose $x^n, y^n \in S^n_\rvx$
and $\eps$ is
a real number satisfying
$0<\eps<<1$. Let
$\calx = [x^n]$ and
$\caly = [y^n]$. Define

\beq
V_a =
\frac{a^{N_\rvx-1}
\pi^{\frac{N_\rvx-1}{2}}}
{\sqrt{N_\rvx}}
\;
\eeq
for any positive real number $a$.
We will
refer to the following
functions
as Dirac delta functions
for setting $\calx$ and
$\caly$ equal

\beq
\delta(\calx, \caly) = \theta(\calx=\caly)
\;,
\eeq

\beq
\delta_\eps(\calx,\caly)=
\exp\left(
-\frac{1}{\eps^2}
\sum_x \{P_\calx(x) - P_\caly(x)\}^2
\right)
\;,
\eeq

\beq
\delta_\eps(x^n,y^n)
= \frac{\delta_\eps(\calx, \caly)}
{\sqrt{d_\calx d_\caly}\;V_{n\eps}}
\;,
\eeq
and

\beq
\delta_\eps(P_\calx-P_\caly)
= \frac{\delta_\eps(\calx, \caly)}{V_\eps}
\;.
\eeq

\begin{claim}
\beq
\sum_{x^n} \delta_\eps(x^n, y^n) = 1
\;,
\eeq
and

\beq
\int \cald P_\calx\;
\delta_\eps(P_\calx - P_\caly) =1
\;.
\eeq
\end{claim}
\proof
This follows from
 integration formula
 Eq.(\ref{eq-p-type-int-over-gauss}).
\qed

\section{Source Coding (Lossy Compression)}

We consider all source coding protocols
that can be
described by the following CB net

\beq
\entrymodifiers={++[o][F-]}
\xymatrix{
\what{\rvx}^n&\rvm\ar[l]&\rvx^n\ar[l]
}
\;,
\eeq
with
$S_\rvx=S_{\what{\rvx}}$ and

\beq
P(x^n) = \prod_{j=1}^n P_\rvx(x_j)
\;,
\eeq

\beq
P(m|x^n) = \delta(m, m(x^n))
\;
\eeq
and

\beq
P(\what{x}^n|m) = \delta(\what{x}^n, \what{x}^n(m))
\;.
\eeq
Assume that we are given
a source $P_\rvx
\in pd(S_\rvx)$.
The encoding function $m(\cdot)$ and
the decoding function $\what{x}^n(\cdot)$
are yet to be specified.\footnote{Many authors
(for instance, C\&T) denote the
encoding function $m(\cdot)$
by $f(\cdot)$
and the decoding function
$\what{x}^n(\cdot)$ by $g(\cdot)$.}

The probability
of error is defined by

\beq
P_{err} = P(\what{\rvx}^n\neq \rvx^n)
\;.
\eeq
We find it more convenient to work
with the probability of success,
which is defined by $P_{suc} =1-P_{err}$.
One has

\beqa
P_{suc} &=& 1 - P_{err}\\
&=&
P(\what{\rvx}^n= \rvx^n)\\
&=&
\sum_{\what{x}^n,m, x^n}
\theta(\what{x}^n=x^n)
P(\what{x}^n|m)P(m|x^n)P_\rvx(x^n)
\\
&=&
\sum_{x^n} P_\rvx(x^n) \delta[x^n,
\what{x}^n\circ m(x^n)]
\;.
\eeqa

Now it's time to
decide what encoding and decoding
functions we want to consider.
Suppose $A$
is a proper subset of $S^n_\rvx$.
One can give each element of $A$
an individual number (its index)
from 1 to $|A|$. Assume, without
loss of generality, that $0^n\not\in A$.
As we shall see, the following
encoding and decoding functions
are good enough:

\beq
m(x^n)
=
\left\{
\begin{array}{ll}
\mbox{index of } x^n \mbox{ in } A
&
\mbox{, if } x^n\in A
\\
0
&
\mbox{, if } x^n\not\in A
\end{array}
\right.
\;,
\eeq
and

\beq
\what{x}^n(m)
=
\left\{
\begin{array}{ll}
m^{-1}(m)
&
\mbox{, if } m\in Z_{1,|A|}
\\
0^n
&
\mbox{, if } m=0
\end{array}
\right.
\;,
\eeq
where the set $A$ is given by
either

\beq
A_{P_\rvx}=\left\{
x^n: R\geq
\frac{1}{n}
\ln \frac{1}{P_\rvx(x^n)}
\right\}
=
\left\{
x^n: R\geq
\sum_x \ptype{x^n}(x)
\ln \frac{1}{P_\rvx(x)}
\right\}
\;,
\eeq
or

\beq
A_{univ} =
\left\{
x^n: R\geq
H(\ptype{x^n})
\right\}
=
\left\{
x^n: R\geq
\sum_x \ptype{x^n}(x)
\ln \frac{1}{\ptype{x^n}(x)}
\right\}
\;
\eeq
for some positive number $R$ yet to be
specified.
These two interesting options
for the set $A$ can be
considered simultaneously
by defining

\beq
A =
\left\{
x^n: R\geq
\sum_x \ptype{x^n}(x)
\ln \frac{1}{Q(x)}
\right\}
\;,
\eeq
where

\beq
Q(x) =
\left\{
\begin{array}{ll}
P_\rvx(x)&\mbox{, source dependent coding}\\
\ptype{x^n}(x)&\mbox{, universal coding}
\end{array}
\;
\right.
\;.
\eeq
In the case of source dependent coding,
$Q$ (and therefore
the functions $m(\cdot)$
and $\what{x}^n(\cdot)$)
depend on the source distribution $P_\rvx$.
In the case of universal coding, $Q$ is
 independent of the source.

Note that for this encoding and decoding
functions,
\beq
\delta[x^n,
\what{x}^n\circ m(x^n)]=
\theta(x^n\in A)=
\theta\left(R\geq
\sum_x \ptype{x^n}(x)
\ln \frac{1}{Q(x)}\right)
\;
\eeq
for all $x^n\in S_\rvx^n-\{0^n\}$
so

\beqa
P_{suc} &=&
\sum_{x^n}
P_\rvx(x^n)
\theta\left(R\geq
\sum_x \ptype{x^n}(x)
\ln \frac{1}{Q(x)}\right)
\\
&\sim&
\int \cald \ptype{x^n}
e^{n \sum_x \ptype{x^n}(x) \ln \frac{P_\rvx(x)}
{\ptype{x^n}(x)}}
\theta\left(R\geq
\sum_x \ptype{x^n}(x)
\ln \frac{1}{Q(x)}\right)
\\
&\approx&
\theta(R\geq H(P_\rvx))
\;.
\label{eq-at-best-q}
\eeqa
Eq.(\ref{eq-at-best-q}) follows
because, as is easily proven,
applying the method
of steepest descent to the p-type integral
yields a tilde point:

\beq
\ptiltype{x^n}(x)=P_\rvx(x)
\;.
\eeq

As mentioned in the notation section,
we define $R_\rvm$ by

\beq
R_\rvm= \frac{\ln N_\rvm}{n}
\;.
\eeq
So far, it's not clear what
value to use for the constant $R$
that appears in the definition of set $A$.
In the next Claim, we will show
that it must equal $R_\rvm$
for our arguments to be valid.

\begin{claim}
\beq
R = R_\rvm
\;
\eeq
for consistency of our arguments.
\end{claim}
\proof
We must have

\beqa
N_\rvm &=&
 \sum_{x^n} \theta(x^n\in A)
 \\
 &\sim&
 \int \cald \ptype{x^n}
 e^{
 n \sum_x\ptype{x^n}(x)\ln
 \frac{1}{\ptype{x^n}(x)}
 }
 \theta\left(R>\sum_x\ptype{x^n}(x)\ln
 \frac{1}{Q(x)}\right)
  \\
 &\sim&
 e^{nR}
 \int \cald \ptype{x^n}
 e^{
 n \sum_x\ptype{x^n}(x)\ln
 \frac{Q(x)}{\ptype{x^n}(x)}
 }
 \theta\left(R>\sum_x\ptype{x^n}(x)\ln
 \frac{1}{Q(x)}\right)
 \\
&\sim&
e^{nR}\theta(R>H(P_\rvx))
\;.
\eeqa
As long as $R>H(\rvx)$,
our approximations are
valid and $N_\rvm  = e^{nR}$.
\qed

\section{Channel Coding}

We define a codebook $\calc$
as an $N_\rvm\times n $ matrix
given by
$\calc = \{x^n(m)\}_{\forall m}= x^n(\cdot)$
where $x^n(m)\in S^n_\rvx$ for all $m\in S_\rvm$.

We consider all
channel coding protocols that can be
described by the following CB net

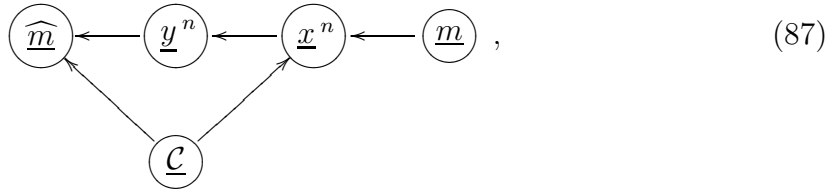
\beq
\entrymodifiers={++[o][F-]}
\xymatrix{
\what{\rvm}&\rvy^n\ar[l]&\rvx^n\ar[l]&\rvm\ar[l]\\
*{}&\rv{\calc}\ar[ul]\ar[ur]&*{}&*{}
}
\;,
\label{eq-ch-qbnet}
\eeq
with

\beq
P(m) = \frac{1}{N_\rvm}
\;,
\eeq

\beq
P(x^n|m, \calc)=
\delta(x^n, x^n(m))
\;,
\eeq

\beq
P(y^n | x^n)
=
\prod_j P(y_j|x_j)
=
e^{n\sum_{x,y} \ptype{x^n,y^n}(x,y)\ln P(y|x)}
\;,
\eeq

\beq
P(\calc)=\mbox{to be specified}
\;,
\eeq
and

\beq
P(\what{m}|y^n,\calc)=\mbox{to be specified}
\;.
\eeq
Assume that we are given
a channel $\{P_{\rvy|\rvx}(y|x)\}_{\forall y}
\in pd(S_\rvy)$
for all $x\in S_\rvx$.
The encoding $P(\calc)$
and
decoding $P(\what{m}|y^n,\calc)$
probability distributions are
yet to be specified.

It's convenient to define
the coding rate $R_\rvm$ by

\beq
R_\rvm = \frac{\ln N_\rvm}{n}
\;
\eeq
and
the channel capacity $C$ by
\beq
C = \max_{P_\rvx} H(\rvy:\rvx)
\;.
\eeq

\begin{claim} \label{cl-ind-bd-mi}(Independence upper bound
for mutual information of DMC)
If $P(y^n|x^n) =\prod_{j=1}^{n}P(y_j|x_j)$ (this
is what is called a discrete memoryless channel, DMC),
then

\beq
H(\rvy^n:\rvx^n)\leq \sum_{j=0}^n
H(\rvy_j:\rvx_j)
\;.
\eeq
Furthermore, equality holds
iff the $\rvx_j$ are
mutually independent.
\end{claim}
\proof
Assume $n=3$ for illustrative purposes.
If the $\rvx_j$
are not independent, we must consider
the following CB net

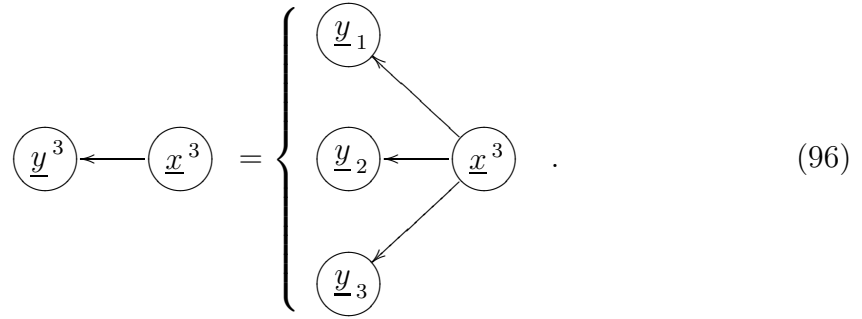
\beq
\begin{array}{c}
\entrymodifiers={++[o][F-]}
\xymatrix{
\rvy^3&\rvx^3\ar[l]
}
\end{array}
=
\left\{
\begin{array}{c}
\entrymodifiers={++[o][F-]}
\xymatrix{
\rvy_1&*{}\\
\rvy_2&\rvx^3\ar[ul]\ar[l]\ar[dl]\\
\rvy_3&*{}
}
\end{array}
\right.
\;.
\label{eq-xj-dependent}
\eeq
If the $\rvx_j$ are independent, then
this becomes
\beq
\begin{array}{c}
\entrymodifiers={++[o][F-]}
\xymatrix{
\rvy^3&\rvx^3\ar[l]
}
\end{array}
=
\left\{
\begin{array}{c}
\entrymodifiers={++[o][F-]}
\xymatrix{
\rvy_1&\rvx_1\ar[l]\\
\rvy_2&\rvx_2\ar[l]\\
\rvy_2&\rvx_3\ar[l]
}
\end{array}
\right.
\;
\eeq

In the case of Eq.(\ref{eq-xj-dependent}),
\beqa
H(\rvy^n:\rvx^n)
&=&
H(\rvy^n) - H(\rvy^n|\rvx^n)=
H(\rvy^n) - \sum_j H(\rvy_j|\rvx_j)
\label{eq-ub-a}
\\
&\leq&
\sum_j H(\rvy_j) - \sum_j H(\rvy_j|\rvx_j)
\label{eq-ub-b}
\\
&=&
\sum_j H(\rvy_j:\rvx_j)
\;
\label{eq-ub-c}
\eeqa
Eq.(\ref{eq-ub-b}) follows from
the ``subadditivity"
or ``independence upper bound"
of the joint entropy, which says that
$H(\rva,\rvb)\leq H(\rva) + H(\rvb)$
for any random variables $\rva$ and $\rvb$.
(See C\&T for a proof of subadditivity).
If the $\rvx_j$ are mutually independent,
then the $\rvy_j$ must be mutually
independent too, in which case
Eq.(\ref{eq-ub-b}) becomes an equality.
Conversely, if
Eq.(\ref{eq-ub-b}) is an equality,
then the $\rvy_j$ must be
mutually independent so
the $\rvx_j$ must be too.
\qed

\begin{claim}
Optimality:
$\forall R_\rvm$, if $\exists$
an encoding and a decoding
that satisfy $\lim_{n\rarrow\infty}P_{err}=0$
for the CB net of
Eq.(\ref{eq-ch-qbnet}),
then $R_\rvm\leq C$.
\end{claim}
\proof
\beqa
nR_\rvm &=& \ln N_\rvm = H(\rvm) =
H(\rvy^n: \rvm) + H(\rvm | \rvy^n)
\label{eq-ch-a}
\\
&\leq&
H(\rvy^n: \rvm) + n\delta
\label{eq-ch-b}
\\
&\leq&
H(\rvy^n: \rvx^n) + n\delta
\label{eq-ch-c}
\\
&\leq&
\sum_{j=1}^n
H(\rvy_j:\rvx_j) + n\delta
\label{eq-ch-d}
\\
&\leq&
n(C + \delta)
\label{eq-ch-e}
\;
\eeqa

\begin{itemize}
\item[(\ref{eq-ch-b}):]
This follows from Fano's inequality.
(See C\&T for a proof
of Fano's inequality.)
$\delta$ is some positive number
that tends to zero as $n\rarrow \infty$

\item[(\ref{eq-ch-c}):] This follows from the data processing
inequalities. (See C\&T for a
proof of the data processing inequalities.)

\item[(\ref{eq-ch-d}):] This
follows from Claim \ref{cl-ind-bd-mi}.

\item[(\ref{eq-ch-e}):] This follows from the definition of
channel capacity $C$.
\end{itemize}
\mbox{\;}
\qed

\begin{claim}
Achievability:
$\forall R_\rvm$,
if $R_\rvm\leq C$, then
$\exists$
an encoding and a decoding
that satisfy $\lim_{n\rarrow\infty}P_{err}=0$
for the CB net of
Eq.(\ref{eq-ch-qbnet}).
\end{claim}
\proof
So far, the
encoding and decoding probability
distributions are unspecified.
In this proof, we will use one
possible choice
for these distributions. This choice,
although not very practical,
turns out to yield optimal results.
For $P(\calc)$ we choose what is
called random coding:

\beq
P(\calc) = P_\rvx(x^n(\cdot))=
\prod_{m}
P_\rvx(x^n(m))=
\prod_{m,j}
P_\rvx(x_j(m))
\;
\eeq
for some source $P_\rvx
\in pd(S_\rvx)$.
For
$P(\what{m}|y^n,\calc)$
we choose a maximum likelihood
decoder:\footnote{By
$\prod_{ m\neq \what{m}}$
we mean $\prod_{ m\in S_\rvm-\{\what{m}\}}$.
}

\beqa
P(\what{m}|y^n,\calc)
&=&
\prod_{ m\neq \what{m}}
\theta\left(
R< \frac{1}{n}\ln
\frac{P(y^n|x^n(\what{m}))}
{P(y^n|x^n(m))}
\right)
\\
&=&
\prod_{ m\neq \what{m}}
\theta\left(
R< \frac{1}{n}\ln
\frac{P(y^n:x^n(\what{m}))}
{P(y^n:x^n(m))}
\right)\;
\eeqa
for some $R>0$.
Note that there is no
guarantee that
this definition of $P(\what{m}|y^n,\calc)$
is a well defined probability distribution
satisfying
$\sum_{\what{m}}P(\what{m}|y^n,\calc)=1$.
In the next Claim, we
will prove
that if $R=R_\rvm$,
then $P(\what{m}|y^n,\calc)$
is well defined.

The probability
of error is defined by

\beq
P_{err} =
P(\what{\rvm}\neq \rvm)
\;.
\eeq
We find it more convenient to work
with the probability of success,
which is defined by $P_{suc} =1-P_{err}$.
One has

\beqa
P_{suc} &=& 1- P_{err}
\\
&=& P(\what{\rvm}=\rvm)
\\
&=&
\sum_{\what{m}, m}
\theta(\what{m}= m) P(\what{m}, m)
\\
&=&
\sum_{\what{m}, m, y^n, x^n, \calc}
\theta(\what{m}=m)
P(\what{m}|y^n, \calc)
P(y^n|x^n)
\delta(x^n, x^n(m))P(m)
P(\calc)
\\
&=&
\frac{1}{N_\rvm}\sum_{\what{m}}
\sum_\calc P(\calc)
\sum_{y^n}
P(\what{m}|y^n, \calc)
P(y^n| x^n(\what{m}))
\;.
\eeqa

The choice of $\what{m}\in S_\rvm$
does not matter. Any choice would
give the same answer for $P_{suc}$
\beq
\frac{1}{N_\rvm}\sum_{\what{m}}
\sum_\calc P(\calc) =
\sum_\calc P(\calc) = E_\calc
\;.
\eeq
Thus

\beq
P_{suc} =
E_\calc
\sum_{y^n}
P(y^n|x^n(\what{m}))
\prod_{m\neq \what{m}}
\theta\left(
R< \frac{1}{n}\ln
\frac{P(y^n:x^n(\what{m}))}
{P(y^n:x^n(m))}
\right)
\;.\label{eq-ch-psuc-pre-int_k}
\eeq

Let

\beq
\oint_{k(\cdot)} =
\prod_{m\neq \what{m}}
\left\{
\int_{-\infty}^{+\infty}
\frac{dk(m)}{2\pi i}\;\;
\frac{1}{(k(m)-i\eps)}
\right\}
\;,
\label{eq-ch-oint-def}
\eeq
and

\beq
K= \sum_{m\neq \what{m}} k(m)
\;.
\eeq

Expressing the $\theta$
functions in Eq.(\ref{eq-ch-psuc-pre-int_k})
 as
integrals (see Eq.(\ref{eq-theta-comp-int})),
we get

\beq
P_{suc} =
\oint_{k(\cdot)}
e^{-iKR}
\sum_{y^n, x^n(\cdot)}
\exp\left(
n \sum_{y\in S_\rvy
\;,\;x(\cdot)\in S_\rvx^{N_\rvm}}
\ptype{\;}(y, x(\cdot))
\ln Z(y, x(\cdot))
\right)
\;,
\eeq
where

\beq
Z(y, x(\cdot))
=
P(y|x(\what{m}))
\prod_m\left\{
P_\rvx(x(m))
\right\}
\prod_{m\neq \what{m}}
\left\{
\frac{P^{i\frac{k(m)}{n}}(y:x(\what{m}))}
{P^{i\frac{k(m)}{n}}(y:x(m))}
\right\}
\;.
\eeq

Next we express the sum over $y^n,x^n(\cdot)$
as a p-type integral to get

\beq
P_{suc} =
\oint_{k(\cdot)}
e^{-iKR}
\int \cald \ptype{\;}
n^{N_\rvy +N_\rvx N_\rvm-1 }
(d_{[y^n, x^n(\cdot)]})_{H=0}
e^{\call_0}
\;,
\label{eq-ch-psuc-pre-p-type-int}
\eeq
where

\beq
\call_0=
n \sum_{y, x(\cdot)}
\ptype{\;}(y, x(\cdot))
\ln \frac{Z(y, x(\cdot))}
{\ptype{\;}(y, x(\cdot))}
\;.
\eeq
We add to $\call_0$ a Lagrange multiplier
term that constrains the components
of the vector
$\{\ptype{\;}(y,x(\cdot))\}_{\forall y, x(\cdot)}$
so that they sum to one:

\beq
\call = \call_\lam =
\call_0 + n\lam \left(
\sum_{y,x(\cdot)}
\ptype{\;}(y,x(\cdot)) -1
\right)
\;
\eeq
for any $\lam\in \RR$.
It's easy to check that $\call$
is maximized when

\beq
\ptiltype{\;}(y, x(\cdot))=
\frac{Z(y, x(\cdot))}
{\sum_{y, x(\cdot)}Z(y, x(\cdot))}
\;.
\eeq
Evaluating the integrand
of the p-type integral
in Eq.(\ref{eq-ch-psuc-pre-p-type-int})
at this tilde point
yields

\beq
P_{suc} =
\oint_{k(\cdot)}
e^{-iKR}
e^{
n\ln Z
}
\;,
\label{eq-ch-psuc-pre-delta}
\eeq
where

\beq
Z =\sum_{y, x(\cdot)}Z(y, x(\cdot))
\;.
\eeq
Using the shorthand notations

\beq
E_y = \sum_y P(y),
\;
E_{x(m)} = \sum_{x(m)} P_\rvx(x(m))
\;,
\eeq
$Z$  can be expressed as

\beq
Z=
E_y
\left[
E_{x(\what{m})}[P^{1+i\frac{K}{n}}(y:x(\what{m}))]
\prod_{m\neq \what{m}}
\left\{
E_{x(m)}[P^{-i\frac{k(m)}{n}}(y:x(m))]
\right\}
\right]
\;.
\label{eq-ch-z-exp}
\eeq

Define
\beq
Z_0= [Z]_{k(m)=0\;\forall m}=
E_y
E_{x(\what{m})}[P^{1+i\frac{K}{n}}(y:x(\what{m}))]
\;.
\label{eq-ch-zo-exp}
\eeq

Note that 1 equals

\beqa
1 &=& \int_{-\infty}^{+\infty}dK\;
\delta(\sum_{m\neq \what{m}}
\left\{k(m)\right\}-K)
\\
&=&
\int_{-\infty}^{+\infty}dK\;
\int_{-\infty}^{+\infty}\frac{dh}{2\pi}\;
e^{ih\left(
\sum_{m\neq \what{m}}\left\{k(m)\right\}-K
\right)}
\;.
\label{eq-ch-one-is}
\eeqa
Multiplying $P_{suc}$
by 1 certainly doesn't change it.
Thus
the right hand sides of
Eqs.(\ref{eq-ch-psuc-pre-delta})
and (\ref{eq-ch-one-is})
can be
multiplied to get

\beq
P_{suc}=
\int_{-\infty}^{+\infty}\frac{dh}{2\pi}\;
\int_{-\infty}^{+\infty}dK\;
e^{iK(-h-R)}
\oint_{k(\cdot)}
e^{ih\sum_{m\neq \what{m}}k(m)}
e^{
n\ln Z
}
\;.
\label{eq-ch-pre-km-int}
\eeq

Next
we will assume that,
for all $m$,
when doing the contour
integration over $k(m)$
in Eq.(\ref{eq-ch-pre-km-int})
with $Z$ given by Eq.(\ref{eq-ch-z-exp}),
the
$e^{n\ln Z}$
can be evaluated at the value
$k(m)=i\eps\rarrow 0$ of the pole.\footnote
{I don't know how to prove this
assumption rigorously.
The assumption is plausible,
and it does lead
to the correct
result for the channel capacity.
It may just be
an approximation that
becomes increasingly good as
$n\rarrow \infty$}
Symbolically, this means we assume

\beqa
\oint_{k(\cdot)}
e^{ih\sum_{m\neq \what{m}}k(m)}
e^{n\ln Z}
&=&
e^{n\ln Z_0}
\oint_{k(\cdot)}
e^{ih\sum_{m\neq \what{m}}k(m)}
\\
&=&
e^{n\ln Z_0}
\theta(h>0)
\;.
\label{eq-ch-magic-contour-int}
\eeqa
Applying Eq.(\ref{eq-ch-magic-contour-int})
to Eq.(\ref{eq-ch-pre-km-int}) gives

\beq
P_{suc}=
\int_{-\infty}^{+\infty}\frac{dh}{2\pi}
\theta(h>0)
\int_{-\infty}^{+\infty}dK\;
e^{iK(-h-R)}
e^{
n\ln Z_0
}
\;.
\label{eq-ch-pre-exp-log}
\eeq

Next we use
Eqs.(\ref{eq-taylor-one})
and (\ref{eq-taylor-two})
to expand $\ln Z_0$ to
second order in $K$. This yields

\beq
\ln Z_0
\approx
i\frac{K}{n} a
-\frac{K^2}{2n^2}b
\;,
\eeq
where

\beq
a= H(\rvy:\rvx)
\;,
\eeq
and

\beqa
b&=&
E_y E_x P(y:x)\ln^2P(y:x)
-H^2(\rvy:\rvx)
\\
&=&
E_{y,x} \ln^2 P(y:x)
-
[E_{y,x}\ln P(y:x)]^2
\\
&\geq& 0
\;
\eeqa
(The inequality follows from
the identity
$\av{\rvx^2}-\av{\rvx}^2 =
\av{(\rvx-\av{\rvx})^2}$
where $\av{\cdot}$
denotes an average
and $\rvx$ is any random variable.)

With the $\ln Z_0$
 expanded to second order in $K$,
Eq.(\ref{eq-ch-pre-exp-log}) becomes

\beq
P_{suc}=
\int_{-\infty}^{+\infty}\frac{dh}{2\pi}\;
\theta(h>0)
\int_{-\infty}^{+\infty}dK\;
e^{iK(a-h-R) -\frac{K^2}{2n}b}
\;.
\eeq
If we keep only the term
linear in $K$ in the argument
of the exponential, we immediately get

\beq
P_{suc}= \theta(R<H(\rvy:\rvx))
\;.
\label{eq-ch-pre-c-limit}
\eeq
If we also keep the term quadratic
in $K$, we get

\beq
P_{suc} = \frac{1}{2}{\rm erfc}\left(
\sqrt{\frac{n}{2b}}[R-H(\rvy:\rvx)]
\right)
\;.
\eeq

Maximizing
both sides of Eq.(\ref{eq-ch-pre-c-limit})
with respect to the source $P_\rvx$,
and using the definition of channel capacity $C$,
we get that there
is an encoding and a decoding
for which

\beq
P_{suc}= \theta(R<C)
\;.
\eeq
\qed

\begin{claim}\label{cl-ch-r-is-rm}
\beq
R = R_\rvm
\;
\eeq
for consistency of our arguments.
\end{claim}
\proof
Rather than
checking that
$\sum_{\what{m}}P(\what{m}|y^n, \calc)=1$,
we will check that
the total probability
distribution for the whole
CB net Eq.(\ref{eq-ch-qbnet})
sums to one.
We want

\beq
1 =
\sum_{\what{m}, m, y^n, x^n, \calc}
P(\what{m}|y^n, \calc)
P(y^n|x^n)
\delta(x^n, x^n(m))P(m)
P(\calc)
\;.
\eeq
Using

\beq
\sum_{\what{m}, m}
=
\sum_{\what{m}, m}
\theta(\what{m}= m)
+
\sum_{\what{m}, m}
\theta(\what{m}\neq m)
\;,
\eeq
and

\beq
\sum_{\what{m}, m}
\theta(\what{m}\neq m)
P(m)\sum_\calc P(\calc)=
\frac{(N_\rvm^2-N_\rvm)}{N_\rvm}
\sum_\calc P(\calc)\approx
N_\rvm E_\calc
\;,
\eeq
we get for
any pair $m_0,\what{m}\in S_\rvm$
such that $m_0\neq\what{m}$,

\beq
1 = P_{suc} + N_\rvm
E_\calc
\sum_{ y^n}
P(\what{m}|y^n, \calc)
P(y^n|x^n(m_0))
\;.
\label{eq-no-dist-one-is-nm}
\eeq
Substituting
into  Eq.(\ref{eq-no-dist-one-is-nm})
the specific values of
the probability distributions
$P(\what{m}|y^n, \calc)$
and $P(y^n|x^n(m_0))$, we get

\beq
P_{err} =  N_\rvm
\int_{-\infty}^{+\infty}\frac{dh}{2\pi}\;
\int_{-\infty}^{+\infty}dK\;
e^{iK(-h-R)}
\oint_{k(\cdot)}
e^{ih\sum_{m\neq \what{m}}k(m)}
e^{
n\ln W
}
\;,
\label{eq-ch-pre-w-to-wo}
\eeq
where
$\oint_{k(\cdot)}$
 is defined as before
(see Eq.(\ref{eq-ch-oint-def})) and
where

\beq
W=
E_y
\left[
\begin{array}{l}
E_{x(\what{m})}[P^{i\frac{K}{n}}(y:x(\what{m}))]
\\
E_{x(m_0)}[P^{1-i\frac{k(m_0)}{n}}(y:x(m_0))]
\\
\prod_{m\neq \what{m}, m_0}
\left\{
E_{x(m)}[P^{-i\frac{k(m)}{n}}(y:x(m))]
\right\}
\end{array}
\right]
\;.
\label{eq-ch-w-def}
\eeq

Let
\beq
W_0 = [W]_{k(m)=0\;\forall m}=
E_y
E_{x(\what{m})}[P^{i\frac{K}{n}}(y:x(\what{m}))]
\;.
\label{eq-ch-wo-def}
\eeq

Next assume that
\beqa
\oint_{k(\cdot)}
e^{ih\sum_{m\neq \what{m}}k(m)}
e^{n\ln W}
&=&
e^{n\ln W_0}
\oint_{k(\cdot)}
e^{ih\sum_{m\neq \what{m}}k(m)}
\\
&=&
e^{n\ln W_0}
\theta(h>0)
\;.
\label{eq-ch-w-to-wo}
\eeqa
Applying Eq.(\ref{eq-ch-w-to-wo})
to Eq.(\ref{eq-ch-pre-w-to-wo}) yields

\beq
P_{err} =  N_\rvm
\int_{-\infty}^{+\infty}\frac{dh}{2\pi}
\theta(h>0)
\int_{-\infty}^{+\infty}dK\;
e^{iK(-h-R)}
e^{
n\ln W_0
}
\;.
\label{eq-ch-wo-pre-shift}
\eeq

Now we can make the following change of variables

\beq
K\rarrow K -in
\;.
\eeq
Note that this change of variables
changes $W_0$ defined by Eq.(\ref{eq-ch-wo-def}) to
$Z_0$ defined by Eq.(\ref{eq-ch-zo-exp}).
Under this change of variables,
Eq.(\ref{eq-ch-wo-pre-shift})
 becomes

\beqa
P_{err}
&=&
N_\rvm
\int_{-\infty}^{+\infty}\frac{dh}{2\pi}
\theta(h>0)
e^{n(-h-R)}
\int_{-\infty}^{+\infty}dK\;
e^{iK(-h-R)}
e^{
n\ln Z_0
}
\\
&\approx &
N_\rvm e^{-nR} P_{suc}
\;,
\eeqa
or,
equivalently,

\beq
\theta(R>H(\rvy:\rvx))\approx
N_\rvm e^{-nR}\theta(R<H(\rvy:\rvx))
\;.
\eeq
Thus, when $R$ equals (or
is very close to) $H(\rvy:\rvx)$,
we must have $N_\rvm  = e^{nR}$.
\qed

\section{Source Coding With Distortion}
Assume that we
are given a function
$d(x,y)$
that measures  the distance
between two letters
of $x,y\in S_\rvx$.
Assume $d(x,x)=0$ and
$d(x,y)\geq 0$
for all $x,y\in S_\rvx$.

Assume that random variables
$\rvx$ and $\what{\rvx}$
both have the same set of possible values $S_\rvx$.
We define codebook
$\calc$
 as an $N_\rvm\times n $ matrix
given by
$\calc = \{x^n(m)\}_{\forall m}= x^n(\cdot)$
where $x^n(m)\in S^n_\rvx$ for all $m\in S_\rvm$.
We define another codebook
$\what{\calc}$
 as an $N_\rvm\times n $ matrix
given by
$\what{\calc} = \{\what{x}^n(m)\}_{\forall m}= \what{x}^n(\cdot)$
where $\what{x}^n(m)\in S^n_\rvx$ for all $m\in S_\rvm$.

We consider all
source coding protocols that can be
described by the following CB net:

\beq
\entrymodifiers={++[o][F-]}
\xymatrix{
\what{\rvx}^n&*{}&\rvm\ar[ll]&\rvx^n\ar[l]\\
*{}&\rv{\what{\calc}}\ar[ul]\ar[ur]&\rv{\calc}\ar[l]&*{}
}
\;
\label{eq-dist-qbnet}
\eeq
with
$S_\rvx=S_{\what{\rvx}}$ and

\beq
P(x^n) = \prod_{j=1}^n P_\rvx(x_j)
\;,
\eeq
\beq
P(m|x^n,\what{\calc})=\mbox{to be specified}
\;,
\eeq

\beq
P(\calc)=\mbox{to be specified}
\;,
\eeq

\beq
P(\what{\calc}|\calc)=
\prod_m P_{\what{\rvx}|\rvx}(\what{x}^n(m)|x^n(m))
=\prod_{m,j} P_{\what{\rvx}|\rvx}(\what{x}_j(m)|x_j(m))
\;,
\eeq
and

\beq
P(\what{x}^n|m,\what{\calc})
= \delta(\what{x}^n, \what{x}^n(m))
\;.
\eeq
Assume that we are given
a source $\{P_\rvx(x)\}_{\forall x}
\in pd(S_\rvx)$ and
a channel $\{P_{\what{\rvx}|\rvx}(\what{x}|x)\}_{\forall \what{x}\in S_\rvx}
\in pd(S_\rvx)$ for all $x\in S_\rvx$.
The encoding $P(m|x^n,\what{\calc})$
and
decoding $P(\calc)$
probability distributions are
yet to be specified.

Henceforth, we will use
the following shorthand notations
\beq
E_j=\frac{1}{n} \sum_{j=1}^n\;,
\;\;
E_{\what{x},x}=\sum_{\what{x},x} P_{\what{\rvx}|\rvx}(\what{x}| x)
P_\rvx(x)
\;.
\eeq

As usual, we define the  {\bf rate of} $\rvm$
by $R_\rvm = \ln(N_\rvm)/n$.
We define the probability of success by
\beq
P_{suc} = P[
E_j d(\what{\rvx}_j, \rvx_j)\leq D]
\;
\label{eq-def-of-d}
\eeq
where $D\in\RR^{>0}$ is called
the {\bf distortion}.
Note that when $D=0$,
$P_{suc}= P(\what{\rvx}^n=\rvx^n)$,
which is what we used
previously when we considered
source coding
without distortion.

For any source $P_\rvx$ and distortion $D$,
it is useful to define a
{\bf rate distortion function}
$H_\rvx(D)$ by

\beq
H_\rvx(D)=
\min_{P_{\what{\rvx}|\rvx}:
E_{\what{x},x} d(\what{x},x)
< D}H_{P_{\what{\rvx}|\rvx}
P_\rvx}(\what{\rvx}:\rvx)
\;.
\eeq

\begin{claim}\label{cl-props-rat-dis}
(Properties of $H_\rvx(D)$)

\begin{itemize}
\item[(a)]
$H_\rvx(D)$ is a monotonically non-increasing,
convex
function of $D$.
\item[(b)]
$H_\rvx(0) = H(\rvx)$
\item[(c)]
$H_\rvx(E^{Q}_{\what{x},x}d(\what{x},x))
\leq H_{Q}(\what{\rvx}:\rvx)$,
where $E^{Q}_{\what{x},x}=\sum_{\what{x},x}
Q(\what{x},x)$,
where $\{Q(\what{x},x)\}_{\forall \what{x},x}
\in pd(S_{\what{\rvx},\rvx})$
such that
$\sum_{\what{x}}Q(\what{x},x)=P_\rvx(x)$
for all $x$.

\end{itemize}
\end{claim}
\proof

proof of $(a)$: Monotonicity is obvious.
To prove convexity, recall
(see C\&T for a proof) that
the mutual information is a convex
function of its joint probability.
This means that
for any $\lam\in [0,1]$
and
$P_1, P_0\in pd(S_{\what{\rvx},\rvx})$,
if

\beq
P_\lam(\what{x},x) =
\lam P_1(\what{x},x) +
(1-\lam)P_0(\what{x},x)
\;
\label{eq-p-lam-def}
\eeq
for all $\what{x},x$,
then

\beq
H_{P_\lam}(\what{\rvx}:\rvx) \leq
\lam H_{P_1}(\what{\rvx}:\rvx) +
(1-\lam)H_{P_0}(\what{\rvx}:\rvx)
\;.
\eeq

For any
$\lam \in [0,1]$,
let
$D_0, D_1\in \RR^{\geq 0}$
and

\beq
D_\lam
=
\lam D_1 + (1-\lam)D_0
\;.
\eeq
Suppose $P_0,P_1\in pd(S_{\what{\rvx},\rvx})$
such that $\sum_{\what{x}}P_j(\what{x},x)=
P_\rvx(x)$
for all $x$ and

\beq
H_\rvx(D_j)=
H_{P_j}(\what{\rvx}:\rvx)
\;
\eeq
for $j=0,1$.
Define $P_\lam$
by Eq.(\ref{eq-p-lam-def}).
Then

\beqa
H_\rvx(D_\lam)&\leq&
H_{P_\lam}(\what{\rvx}:\rvx)
\\
&\leq&
\lam H_{P_1}(\what{\rvx}:\rvx) +
(1-\lam)H_{P_0}(\what{\rvx}:\rvx)
\\
&=&
\lam H_\rvx(D_1) +
(1-\lam)H_\rvx(D_0)
\;.
\eeqa

proof of $(b)$: If $D=0$, then
$P(\what{x}|x)=\delta_x^{\what{x}}$
so $H(\what{\rvx}:\rvx) = H(\rvx)$.

proof of $(c)$: This follows from definition of
$H_\rvx(D)$.
\qed

\begin{claim} Optimality:
$\forall (D, R_\rvm)$, if $\exists$
an encoding and a decoding
that satisfy $\lim_{n\rarrow\infty}P_{err}=0$
for the CB net
of Eq.(\ref{eq-dist-qbnet}), then
 $R_\rvm\geq H_\rvx(D)$.
\end{claim}
\proof

\beqa
nR_\rvm &=& \ln N_\rvm = H(\rvm) =
H(\what{\rvx}^n:\rvm) + H(\rvm|\what{\rvx}^n)
\label{eq-dis-a}
\\
&\geq&
H(\what{\rvx}^n:\rvm)
\label{eq-dis-b}
\\
&\geq&
H(\what{\rvx}^n:\rvx^n)
\label{eq-dis-c}
\\
&=&
\sum_j H(\what{\rvx}_j: \rvx_j)
\label{eq-dis-d}
\\
&\geq&
\sum_j H_{\rvx}
\left(E_{\what{x}_j,x_j}d(\what{x}_j,x_j)\right)
\label{eq-dis-e}
\\
&\geq&
 n H_{\rvx}
\left(\frac{1}{n}
\sum_j E_{\what{x}_j,x_j}d(\what{x}_j,x_j)\right)
\label{eq-dis-f}
\\
&=&
n H_{\rvx}
\left(
E_{\what{x},x}d(\what{x},x)\right)
\label{eq-dis-g}
\\
&\geq&
nH_\rvx(D)
\label{eq-dis-h}
\;
\eeqa
\begin{itemize}

\item[(\ref{eq-dis-c}):]
This follows from the data processing
inequalities. (See C\&T for a
proof of the data processing inequalities.)

\item[(\ref{eq-dis-d}):]
This
follows from Claim \ref{cl-ind-bd-mi}
in the case of equality.
We are assuming that
$P(\what{\calc}|\calc)$
is a DMC, and that $P(\calc)$
is an i.i.d. source. This forces
$(\what{x}_j(m), x_j(m))$
and
$(\what{x}_{j'}(m), x_{j'}(m))$
with $j\neq j'$
to be independent.

\item[(\ref{eq-dis-e}):]
This follows from Claim \ref{cl-props-rat-dis},
part (c).
\item[(\ref{eq-dis-f}):]
This follows because $H_\rvx(D)$
is a convex function of $D$.
\item[(\ref{eq-dis-g}):]
This follows
from
using $\ptype{\;}(\what{x},x)\rarrow
P(\what{x},x)$.

\item[(\ref{eq-dis-h}):]
 Eq.(\ref{eq-def-of-d})
 is the definition
of $D$. Expressing
Eq.(\ref{eq-def-of-d})
in terms of p-types
and using $\ptype{\;}(\what{x},x)\rarrow
P(\what{x},x)$,
 we find that
$E_{\what{x},x} d(\what{x},x)<D$
is necessary for success.
Then use the fact that
$H_\rvx(D)$ is non-increasing.
\end{itemize}
\mbox{\;}
\qed

\begin{claim} Achievability:
$\forall (D, R_\rvm)$,
if $R_\rvm\geq H_\rvx(D)$, then
$\exists$
an encoding and a decoding
that satisfy $\lim_{n\rarrow\infty}P_{err}=0$
for the CB net
of Eq.(\ref{eq-dist-qbnet}).
\end{claim}
\proof
So far, the
encoding and decoding probability
distributions are unspecified.
In this proof, we will use one
possible choice
for these distributions.
For decoder $P(\calc)$ we choose:

\beq
P(\calc)
=
P_\rvx(x^n(\cdot))
=
\prod_m
\left\{
P_\rvx(x^n(m))
\right\}
=
\prod_{m,j}
\left\{
P_\rvx(x_j(m))
\right\}
\;,
\eeq
and for encoder
$P(m|x^n, \what{\calc})$
we choose:

\beqa
P(m|x^n, \what{\calc})
&=&
\prod_{m'\neq m}
\theta\left(
R> \frac{1}{n}
\ln
\frac{P(x^n|\what{x}^n(m))}
{P(x^n|\what{x}^n(m'))}
\right)
\\
&=&
\prod_{m'\neq m}
\theta\left(
R> \frac{1}{n}
\ln
\frac{P(x^n:\what{x}^n(m))}
{P(x^n:\what{x}^n(m'))}
\right)
\;
\label{eq-dist-pm}
\eeqa
for some $R>0$.
Note that there is no
guarantee that
this definition of $P(m|x^n, \what{\calc})$
is a well defined probability distribution
satisfying
$\sum_{m}P(m|x^n, \what{\calc})=1$.
In the next Claim, we
will prove
that if $R=R_\rvm$,
then $P(m|x^n, \what{\calc})$
is well defined.

Let
\beq
P(\what{\calc}) =
 \sum_\calc P(\what{\calc}|\calc)P(\calc)
\;.
\eeq

One has
\beqa
P_{suc} &=&
P[ E_j d(\what{\rvx}_j, \rvx_j)<D]
\\
&=&
\sum_{\what{x}^n, x^n}
P(\what{x}^n, x^n)
\theta(E_j d(\what{x}_j, x_j)<D)
\\
&=&
\sum_{\what{x}^n, x^n, m, \what{\calc}}
P(\what{x}^n|m, \what{\calc})
P(m|x^n, \what{\calc})
P(x^n)P(\what{\calc})
\theta(E_j d(\what{x}_j, x_j)<D)
\\
&=&
\sum_m E_{\what{\calc}}
E_{x^n}
P(m|x^n, \what{\calc})
\theta(E_j d(\what{x}_j(m), x_j)<D)
\;.
\label{eq-dist-pre-m-is-one}
\eeqa

Consider what happens to
$P(m|x^n, \what{\calc})$
in
Eq.(\ref{eq-dist-pre-m-is-one})
as
$D\rarrow 0$.
When $D\rarrow 0$,
$\what{x}^n(m)\rarrow x^n$
by virtue of Eq.(\ref{eq-dist-pre-m-is-one}).
Hence
$P(x^n|\what{x}^n(m))\rarrow 1$.
Furthermore,
$P(x^n|\what{x}^n(m'))\rarrow
P(x^n(m)|\what{x}^n(m'))= P(x^n(m))\delta_m^{m'}
=P(x^n)\delta_m^{m'}$.
Thus

\beq
P(m|x^n, \what{\calc})
\rarrow
\theta\left(
R> \frac{1}{n}
\ln
\frac{1}
{P(x^n)}
\right)
= \theta(x^n\in A_{P_\rvx})
\;.
\eeq
Hence, when $D=0$, the encoder
$P(m|x^n, \what{\calc})$
in
Eq.(\ref{eq-dist-pre-m-is-one})
is the same as the one we
used when we considered
source coding without distortion.

For any $Q\in pd(S_{\what{\rvx},\rvx})$
such that $\sum_{\what{x}}Q(\what{x},x)=P_\rvx(x)$
for all $x$, define
\beq
\theta_{Q(\what{x},x)}
=
\theta_{Q_{\what{\rvx},\rvx}}
=
\theta(\sum_{\what{x},x}Q(\what{x},x)d(\what{x},x)<D)
\;.
\eeq
Note that

\beq
\theta(E_j d(\what{x}_j(1), x_j)<D)
=
\theta_{\ptype{\;}(\what{x}(1), x)}
\;.
\label{eq-theta-j-to-ptype}
\eeq

Note that
\beq
\sum_m E_{\what{\calc}}=
N_\rvm E_{\what{\calc}}
\;.
\eeq
Hence, the choice of $m\in S_\rvm$
in Eq.(\ref{eq-dist-pre-m-is-one})
does not matter. Any choice would
give the same answer for $P_{suc}$.
Thus,
Eq.(\ref{eq-dist-pre-m-is-one})
can be replaced by
the following.
Assume
$1\in S_\rvm$
and replace
$m$ by 1 and $m'$ by $m$.
Also use
Eq.(\ref{eq-theta-j-to-ptype}).
Then

\beqa
P_{suc}&=&
N_\rvm
E_{\what{\calc}}
E_{x^n}
\prod_{m\neq 1}
\left\{
\theta\left(
R> \frac{1}{n}
\ln
\frac{P(x^n:\what{x}^n(1))}
{P(x^n:\what{x}^n(m))}
\right)\right\}
\theta_{\ptype{\;}(\what{x}(1), x)}
\;.
\label{eq-dist-psuc-pre-int_k}
\eeqa

If we assume that
our formalism will eventually
justify the physically plausible assumption
that
$\ptype{\;}(\what{x}(1), x)\rarrow
P_{\what{\rvx},\rvx}(\what{x}(1), x)$,
then we may replace
$\theta_{\ptype{\;}(\what{x}(1), x)}$
by $\theta_{P_{\what{\rvx}, \rvx}}$
at this point. This would
simplify the analysis below.
Instead, we will
continue with
$\theta_{\ptype{\;}(\what{x}(1), x)}$
and show that
our formalism does indeed
lead to the same result
as if we had
replaced
$\theta_{\ptype{\;}(\what{x}(1), x)}$
by $\theta_{P_{\what{\rvx}, \rvx}}$ at this point.

Let
\beq
\oint_{k(\cdot)} =
\prod_{m\neq 1}
\left\{
\int_{-\infty}^{+\infty}
\frac{dk(m)}{2\pi i}\;\;
\frac{1}{(k(m)-i\eps)}
\right\}
\;,
\eeq
and

\beq
K= \sum_{m\neq 1} k(m)
\;.
\eeq

Expressing the $\theta$
functions in Eq.(\ref{eq-dist-psuc-pre-int_k})
 as
integrals (see Eq.(\ref{eq-theta-comp-int})),
we get

\beq
P_{suc} = N_\rvm
\oint_{k(\cdot)}
e^{iKR}
\sum_{\what{x}^n(\cdot), x^n}
\exp\left(
n \sum_{\what{x}(\cdot)\in S_\rvx^{N_\rvm}
\;,\; x\in S_\rvx}
\ptype{\;}(\what{x}(\cdot), x)
\ln Z(\what{x}(\cdot), x)
\right)
\theta_{\ptype{\;}(\what{x}(1), x)}
\;,
\eeq
where

\beq
Z(\what{x}(\cdot), x)
=
P(x)
\prod_m\left\{
P(\what{x}(m))
\right\}
\prod_{m\neq 1}
\left\{
\frac{P^{-i\frac{k(m)}{n}}(x:\what{x}(1))}
{P^{-i\frac{k(m)}{n}}(x:\what{x}(m))}
\right\}
\;.
\eeq

Next we express the sum over $\what{x}^n(\cdot), x^n$
as a p-type integral to get

\beq
P_{suc} = N_\rvm
\oint_{k(\cdot)}
e^{iKR}
\int \cald \ptype{\;}
n^{N_\rvx(N_\rvm+1)-1}
(d_{[\what{x}^n(\cdot), x^n]})_{H=0}
e^{\call_0}
\theta_{\ptype{\;}(\what{x}(1), x)}
\;,
\label{eq-dist-psuc-pre-doing-p-type-int}
\eeq
where

\beq
\call_0=
n \sum_{\what{x}(\cdot), x}
\ptype{\;}(\what{x}(\cdot), x)
\ln \frac{Z(\what{x}(\cdot), x)}
{\ptype{\;}(\what{x}(\cdot), x)}
\;.
\eeq
We add to $\call_0$ a Lagrange multiplier
term that constrains the components
of the vector
$\{\ptype{\;}(\what{x}(\cdot), x)\}_{\forall \what{x}(\cdot), x}$
so that they sum to one:

\beq
\call = \call_\lam =
\call_0 + n\lam \left(
\sum_{\what{x}(\cdot), x}
\ptype{\;}(\what{x}(\cdot), x) -1
\right)
\;
\eeq
for any $\lam\in \RR$.
It's easy to check that $\call$
is maximized when

\beq
\ptiltype{\;}(\what{x}(\cdot), x)=
\frac{Z(\what{x}(\cdot), x)}
{\sum_{\what{x}(\cdot), x}Z(\what{x}(\cdot), x)}
\;.
\eeq
Evaluating the integrand
of the p-type integral
in Eq.(\ref{eq-dist-psuc-pre-doing-p-type-int})
at this tilde point
yields

\beq
P_{suc} =  N_\rvm
\oint_{k(\cdot)}
e^{iKR}
e^{
n\ln Z
}\theta_{\ptiltype{\;}(\what{x}(1), x)}
\label{eq-dist-psuc-pre-delta}
\eeq
where

\beq
Z =\sum_{\what{x}(\cdot), x}Z(\what{x}(\cdot), x)
\;.
\eeq

$Z$  can be expressed as

\beq
Z=
E_x
\left[
E_{\what{x}(1)}[P^{-i\frac{K}{n}}(\what{x}(1):x)]
\prod_{m\neq 1}
\left\{
E_{\what{x}(m)}[P^{i\frac{k(m)}{n}}(\what{x}(m):x)]
\right\}
\right]
\;.
\label{eq-dist-z-exp}
\eeq

Define

\beq
Z_0=
[Z]_{k(m)=0\;\forall m}
=
E_x
E_{\what{x}(1)}[P^{-i\frac{K}{n}}(\what{x}(1):x)]
\;.
\label{eq-dist-zo-exp}
\eeq

Note that 1 equals

\beqa
1 &=& \int_{-\infty}^{+\infty}dK\;
\delta(\sum_{m\neq 1}
\left\{k(m)\right\}-K)
\\
&=&
\int_{-\infty}^{+\infty}dK\;
\int_{-\infty}^{+\infty}\frac{dh}{2\pi}\;
e^{ih\left(
\sum_{m\neq 1}\left\{k(m)\right\}-K
\right)}
\;.
\label{eq-dist-one-is}
\eeqa
Multiplying $P_{suc}$
by 1 certainly doesn't change it.
Thus
the right hand sides of
Eqs.(\ref{eq-dist-psuc-pre-delta})
and (\ref{eq-dist-one-is})
can be
multiplied to get

\beq
P_{suc}=  N_\rvm
\int_{-\infty}^{+\infty}\frac{dh}{2\pi}\;
\int_{-\infty}^{+\infty}dK\;
e^{iK(-h+R)}
\oint_{k(\cdot)}
e^{ih\sum_{m\neq 1}k(m)}
e^{
n\ln Z
}
\theta_{\ptiltype{\;}(\what{x}(1), x)}
\;.
\label{eq-dist-pre-km-int}
\eeq
Next
we will assume that,
for all $m$,
when doing the contour
integration over $k(m)$
in Eq.(\ref{eq-dist-pre-km-int})
with $Z$ given by Eq.(\ref{eq-dist-z-exp}),
the
$e^{n\ln Z}
\theta_{\ptiltype{\;}(\what{x}(1), x)}$
can be evaluated at the value
$k(m)=i\eps\rarrow 0$ of the pole.\footnote
{I don't know how to prove this
assumption rigorously.
The assumption is plausible,
and it does lead
to the correct
result for the channel capacity.
It may just be
an approximation that
becomes increasingly good as
$n\rarrow \infty$}
Symbolically, this means we assume

\beqa
\oint_{k(\cdot)}
e^{ih\sum_{m\neq 1}k(m)}
e^{n\ln Z}
\theta_{\ptiltype{\;}(\what{x}(1), x)}
&=&
e^{n\ln Z_0}
\theta_{P^{-i\frac{K}{n}}(\what{x}(1), x)}
\oint_{k(\cdot)}
e^{ih\sum_{m\neq 1}k(m)}
\\
&=&
e^{n\ln Z_0}
\theta_{P^{-i\frac{K}{n}}(\what{x}(1), x)}
\theta(h>0)
\;.
\label{eq-dist-magic-contour-int}
\eeqa
Applying Eq.(\ref{eq-dist-magic-contour-int})
to Eq.(\ref{eq-dist-pre-km-int}) gives

\beq
P_{suc}=
N_\rvm
\int_{-\infty}^{+\infty}\frac{dh}{2\pi}
\theta(h>0)
\int_{-\infty}^{+\infty}dK\;
e^{iK(-h+R)}
e^{
n\ln Z_0
}
\theta_{P^{-i\frac{K}{n}}(\what{x}(1), x)}
\;.
\label{eq-dist-post-km-int}
\eeq

Next we make the following
change of variables:

\beq
K\rarrow K + in
\;.
\eeq
Let

\beq
W_0=
[Z_0]_{K\rarrow K + in}
=
E_x
E_{\what{x}(1)}[P^{1-i\frac{K}{n}}(\what{x}(1):x)]
\;.
\label{eq-dist-wo-exp}
\eeq
Under this change of variables,
Eq.(\ref{eq-dist-post-km-int}) becomes

\beq
P_{suc}=
N_\rvm
\int_{-\infty}^{+\infty}\frac{dh}{2\pi}
\theta(h>0)
e^{-n(-h+R)}
\int_{-\infty}^{+\infty}dK\;
e^{iK(-h+R)}
e^{
n\ln W_0
}
\theta_{P^{1-i\frac{K}{n}}(\what{x}(1), x)}
\;.
\label{eq-dist-pre-exp-log}
\eeq

Next we use
Eqs.(\ref{eq-taylor-one})
and (\ref{eq-taylor-two})
to expand $\ln W_0$ to
second order in $K$. This yields

\beq
\ln W_0
\approx
-i\frac{K}{n} a
-\frac{K^2}{2n^2}b
\;,
\eeq
where

\beq
a= H(\what{\rvx}:\rvx)
\;,
\eeq
and

\beqa
b&=&
E_{\what{x}} E_x P(\what{x}:x)\ln^2P(\what{x}:x)
-H^2(\what{\rvx}:\rvx)
\\
&=&
E_{\what{x},x} \ln^2 P(\what{x}:x)
-
[E_{\what{x},x}\ln P(\what{x}:x)]^2
\\
&\geq& 0
\;.
\eeqa

With the $\ln W_0$
 expanded to second order in $K$,
 and $\theta_{P^{1-i\frac{K}{n}}(\what{x}(1), x)}$
 to zeroth order in $K$,
Eq.(\ref{eq-dist-pre-exp-log}) becomes

\beq
P_{suc}=
\theta_{P_{\what{\rvx}, \rvx}} N_\rvm
\int_{-\infty}^{+\infty}\frac{dh}{2\pi}\;
\theta(h>0)
e^{n(h-R)}
\int_{-\infty}^{+\infty}dK\;
e^{iK(-a-h+R) -\frac{K^2}{2n}b}
\;.
\eeq
If we keep only the term
linear in $K$ in the argument
of the exponential, we immediately get

\beq
P_{suc}\approx \theta_{P_{\what{\rvx}, \rvx}}
N_\rvm e^{-na}\theta(R>a)
\approx
N_\rvm e^{-nR}
\theta(R>H(\what{\rvx}:\rvx))
\;.
\label{eq-dist-pre-rate-dist-lim}
\eeq

Minimizing both sides of
Eq.(\ref{eq-dist-pre-rate-dist-lim})
with respect to the channel $P_{\what{x}|x}$
and using
the definition of the rate
distortion function $H_\rvx(D)$,
we get
that there is an encoding and a decoding
for which

\beq
P_{suc}= N_\rvm e^{-nR}\theta(R> H_\rvx(D))
\;.
\label{eq-dist-post-rate-dist-lim}
\eeq
\qed

\begin{claim}
\beq
R = R_\rvm
\;
\eeq
for consistency of our arguments.
\end{claim}
\proof
For consistency, must have
$N_\rvm e^{-nR}=1$
in Eq.(\ref{eq-dist-post-rate-dist-lim}).
\qed

\appendix
\section{Appendix: Some Integrals Over Polytopes}

This appendix
is a collection of
integration formulas
for doing
integrals
over polytope shaped regions. These
formulas are
useful for doing p-type integrations.

The standard polytope
is defined as the
set
$\Delta^n = \{(t_0, t_1, \ldots, t_n):
t_0 + t_1 + \ldots +t_n = 1, t_j\geq 0\mbox{ for all } j\}$.

For $\{P_x\}_{\forall x} \in pd(S_\rvx)$,
we define the following  integration operator:

\beq
\int \cald P_\rvx
=
\prod_x
\left\{
\int_0^1 dP_x
\right\}
\delta\left(\sum_x P_x -1\right)
\;.
\eeq
This is the same definition as
Eq.(\ref{eq-p-type-int-measure}),
except for an
arbitrary vector
$\{P_x\}_{\forall x}$
instead of just
for a p-type
$\{\ptype{x^n}(x)\}_{\forall x}$.

It is well known and easy to show
by induction that
\beq
\int \cald P_\rvx\;1 =
\frac{1}{(N_\rvx-1)!}
\;.
\label{eq-caldpx-one}
\eeq
More generally, the
so called Dirichlet integral,
defined by

\beqa
I_n &=&
\prod_{j=1}^n
\left\{
\int_0^1 dx_j\;\; x_j^{a_j-1}
\right\} \int_0^1 dx_0
\delta\left(\sum_{j=0}^n x_j -1\right)
\\
&=&
\prod_{j=1}^n
\left\{
\int_0^1 dx_j\;\; x_j^{a_j-1}
\right\}
\theta(\sum_{j=1}^n x_j \leq 1)
\;
\eeqa
can be shown\footnote{See, for example, Ref.\cite{Jef}
for a proof.} to be equal to

\beq
I_n=
\frac{\prod_{j=1}^n \Gamma(a_j)}
{ \Gamma(\sum_{j=1}^na_j)}
\;,
\eeq
where $\Gamma(\cdot)$ stands
for the Gamma function. $\Gamma(n) = (n-1)!$
for any positive integer $n$.

In SIT, when doing p-type
integrals for large $n$, one often
encounters integrals
of sharply peaked Gaussian
functions integrated over
polytope regions. Since
the Gaussians are
sharply peaked, as long as
their peak is not near the
boundary of the polytope
region, the integrals
can be easily evaluated
approximately
in a Gaussian approximation
which becomes increasingly accurate
as $n$ increases.

Recall that
\beq
\int_{-\infty}^{+\infty} dx\; e^{-\lam x^2} =
\sqrt{\frac{\pi}{\lam}}
\;
\eeq
for $\lam>0$.

\begin{claim}
Suppose $\{Q_x\}_{\forall x} \in pd(S_\rvx)$,
$\Delta P_x = P_x -Q_x$,
and $\lam_x >>1$ for all $x\in S_\rvx$. Then
\beq
\int \cald P_\rvx\;\;
\exp\left(-\sum_x\lam_x(\Delta P_x)^2\right)
\approx
\sqrt{
\frac{\pi^{N_\rvx-1}}
{
\prod_x\left\{\lam_x\right\}
\left(\frac{1}{\lam_\parallel}\right)
}
}
\;,\label{eq-p-type-int-over-gauss}
\eeq
where
$\lam_\parallel =
\left(\sum_x \frac{1}{\lam_x}\right)^{-1}$.
(If the $\lam_x$ are thought of as electrical
resistances connected in parallel,
then $\lam_\parallel$ is
the equivalent resistance.)
\end{claim}
\proof
Let LHS and RHS denote
the left hand side and right hand side
of Eq.(\ref{eq-p-type-int-over-gauss}).
One has

\beqa
LHS &\approx&
\prod_x
\left\{
\int_{-\infty}^{+\infty}
d\Delta P_x
\right\}
\delta(\sum_x\Delta P_x)
\exp
\left(
-\sum_x \lam_x (\Delta P_x)^2
\right)
\\
&=&
\int_{-\infty}^{+\infty}
\frac{dk}{2\pi}\;
\Gamma
\;
\eeqa
where

\beqa
\Gamma
&=&
\prod_x
\left\{
\int_{-\infty}^{+\infty}
d \Delta P_x\;
\exp\left(
-
\lam_x (\Delta P_x)^2 + ik\Delta P_x
\right)
\right\}
\\
&=&
\prod_x
\left\{
e^{-\frac{k^2}{4\lam_x}}
\int_{-\infty}^{+\infty}
d \Delta P_x
\exp\left(
-
\lam_x (\Delta P_x - \frac{ik}{2\lam_x})^2
\right)
\right\}
\\
&=&
e^{-\frac{k^2}{4\lam_\parallel}}
\prod_x\left\{
\sqrt{\frac{\pi}{\lam_x}}
\right\}
\;.
\eeqa
Thus

\beqa
LHS &=& \prod_x \left\{
\sqrt{\frac{\pi}{\lam_x}}
\right\}
\int_{-\infty}^{+\infty}
\frac{dk}{2\pi}\;
e^{-\frac{k^2}{4\lam_\parallel}}
\\&=&
\prod_x \left\{
\sqrt{\frac{\pi}{\lam_x}}
\right\}
\frac{1}{2\pi}
\sqrt{\frac{\pi}{\frac{1}{4\lam_\parallel}}}
\\
&=&
 RHS
\;.
\eeqa
\qed

\begin{claim}
Suppose matrix $(A_{x,x'})_{\forall x,x'}$
has eigenvalues $\{\lam_x\}_{\forall x}$.
Suppose $\{Q_x\}_{\forall x} \in pd(S_\rvx)$,
$\Delta P_x = P_x -Q_x$,
and $\lam_x >>1$ for all $x\in S_\rvx$.
Then
\beq
\int \cald P_\rvx\;\;
\exp\left(-\sum_{x,x'}
\Delta P_x A_{x,x'} \Delta P_{x'}\right)
\approx
\sqrt{
\frac{\pi^{N_\rvx-1}}
{
\det(A) \tr(A^{-1})
}
}
\;,
\eeq
\end{claim}
\proof
Just diagonalize the matrix $A_{x,x'}$ and
use the previous claim, where now the $\lam_x$ are
the eigenvalues of $A$.
\qed

For $\{P_{y|x}\}_{\forall y} \in pd(S_\rvy)$
for all $x\in S_\rvx$, we define
the following  integration operator:

\beq
\int \cald P_{\rvy|\rvx}=
\prod_{x,y}
\left\{ \int_0^1 dP_{y|x}\right\}
\prod_x
\left\{\delta\left(\sum_y P_{y|x}-1\right)
\right\}
\;.
\eeq
This is the same definition as
Eq.(\ref{eq-cond-p-type-int-measure}),
except for an
arbitrary vector
$\{P_{y|x}(y|x)\}_{\forall y}$
instead of just
for a p-type
$\{\ptype{y^n|x^n}(y|x)\}_{\forall y}$.

Note that Eq.(\ref{eq-caldpx-one})
implies that

\beq
\int \cald P_{\rvy|\rvx}\;\;1=
\left[
\frac{1}
{(N_\rvy-1)!}
\right]^{N_\rvx}
\;.
\eeq

\begin{claim}
Suppose matrix $A_{y|x\:,\: y'|x'}$
has eigenvalues $\{\lam_{y|x}\}_{\forall x,y}$.
Suppose $\{Q_{y|x}\}_{\forall y} \in pd(S_\rvy)$,
$\Delta P_{y|x} = P_{y|x} -Q_{y|x}$,
and $\lam_{y|x} >>1$ for all $x\in S_\rvx$
and $y\in S_\rvy$.
Then
(using Einstein's repeated index summation
convention)

\beq
\int \cald P_{\rvy|\rvx}\;\;
\exp\left(-\Delta P_{y|x}
A_{y|x\:,\:y'|x'}
\Delta P_{y'|x'}\right)
\approx
\sqrt{
\frac{
\pi^{N_\rvy N_\rvx-N_\rvx}
}
{
\det(A)\det
\left[\left(\sum_{y_1,y_2}
A^{-1}_{y_1|x_1\:,\:y_2|x_2}\right)_{\forall x_1,x_2}
\right]
}
}
\;,
\label{eq-cond-p-type-int-over-gauss}
\eeq
\end{claim}
\proof
Let LHS and RHS denote
the left hand side and right hand side
of Eq.(\ref{eq-cond-p-type-int-over-gauss}).
Let $(\omega_y)_{y\in S_\rvy}$
be a vector with all components
equal to one.
Then

\beqa
LHS &\approx &
\prod_{x,y}
\left\{
\int_{-\infty}^{+\infty}
d\Delta P_{y|x}
\right\}
\prod_x
\left\{
\delta(\omega_y \Delta P_{y|x})
\right\}
e^{-\Delta P_{y|x}
A_{y|x\:,\:y'|x'}
\Delta P_{y'|x'}}
\\
&=&
\prod_x
\left\{
\int_{-\infty}^{+\infty}
\frac{dk_x}{2\pi}
\right\}
\Gamma
\;,
\eeqa
where

\beqa
\Gamma &=&
\prod_{x,y}
\left\{
\int_{-\infty}^{+\infty}
d\Delta P_{y|x}
\right\}
e^{-\Delta P_{y|x}
A_{y|x\:,\:y'|x'}
\Delta P_{y'|x'}
+ i \omega_y \Delta P_{y|x} k_x
}
\\
&=&
e^{-\frac{1}{4}
k_{x_1}\omega_{y_1}
A^{-1}_{y_1|x_1\:,\:y_2|x_2}
\omega_{y_2}k_{x_2}
}
\prod_{x,y}
\left\{
\int_{-\infty}^{+\infty}
d\Delta P_{y|x}
\right\}
e^{-\widetilde{\Delta} P_{y|x}
A_{y|x\:,\:y'|x'}
\widetilde{\Delta} P_{y'|x'}
}
\;,
\eeqa
where

\beq
\widetilde{\Delta} P_{y|x} =
\Delta P_{y|x} -
\frac{i}{2}
k_{x_1} \omega_{y_1} A^{-1}_{y_1|x_1\:,\:y|x}
\;.
\eeq
Thus

\beq
\Gamma =
e^{-\frac{1}{4}
k_{x_1}\omega_{y_1}
A^{-1}_{y_1|x_1\:,\:y_2|x_2}
\omega_{y_2}k_{x_2}
}
\sqrt{
\frac{\pi^{N_\rvx N_\rvy}}{\det A}
}
\;.
\eeq
Thus

\beqa
LHS &=&
\sqrt{
\frac{\pi^{N_\rvx N_\rvy}}{\det A}
}
\prod_x
\left\{
\int_{-\infty}^{+\infty}
\frac{dk_x}{2\pi}
\right\}
e^{-\frac{1}{4}
k_{x_1}\omega_{y_1}
A^{-1}_{y_1|x_1\:,\:y_2|x_2}
\omega_{y_2}k_{x_2}
}
\\
&=&
\sqrt{
\frac{\pi^{N_\rvx N_\rvy}}{\det A}
}
\frac{\pi^{\frac{N_\rvx}{2}}}
{(2\pi)^{N_\rvx}}
\frac{1}
{\sqrt{\det\left[\left(
\frac{
\omega_{y_1} A^{-1}_{y_1|x_1\:,\:y_2|x_2}
\omega_{y_2}}
{4}
\right)_{\forall x_1,x_2}\right]
}
}
\\
&=& RHS
\;.
\eeqa
\qed

When using many of the integration
formulas presented
in this appendix,
it is necessary
to calculate the
inverse and determinant
of a large matrix.
I found the following
formulas can often
be helpful
in doing this.

\begin{claim}
Suppose
$E$ is an $n\times n$ matrix.
Suppose $p$ and $q$ are $n$ component
column vectors. Suppose

\beq
A = E + pq^T
\;.\label{eq-def-a-is-e-pq}
\eeq
Then
\begin{subequations}
\beq
A^{-1} = E^{-1} -
\frac{ E^{-1} p q^T E^{-1}}
{1 + q^T E^{-1} p }
\;,\label{eq-inv-o-sum}
\eeq

\beq
\det(A) =
\det(E)
(1 + q^T E^{-1} p)
\;.\label{eq-det-o-sum}
\eeq
\end{subequations}

\end{claim}
\proof
To prove Eq.(\ref{eq-inv-o-sum}),
just show that the right
hand sides of Eqs.(\ref{eq-def-a-is-e-pq})
and (\ref{eq-inv-o-sum}) multiply to one.

To prove Eq.(\ref{eq-det-o-sum}),
one may proceed as follows.
We will
assume $A\in \CC^{3\times 3}$
for concreteness.
The proof we will give generalizes
easily to $A$'s of
dimension different from 3.
Let $\eps_{j_1 j_2, j_3}$
be the totally antisymmetric tensor
with 3 indices. We will
use Einstein summation convention.
Let

\beq
Q_j = q_k (E^{-1})_{k,j}
\;.
\eeq
Then

\beqa
\det(A) &=&
\det(E)\det( \delta_{i,j} +
p_i Q_j)
\\
&=&
\det(E)\eps_{j_1,j_2,j_3}
(\delta_{1,j_1} + p_1 Q_{j_1})
(\delta_{2,j_2} + p_2 Q_{j_2})
(\delta_{3,j_3} + p_3 Q_{j_3})
\\
&=&
\det(E)(1 + p_j Q_j)
\;.
\eeqa
\qed

\begin{claim}
Suppose
$A$ is an $n\times n$ matrix, and $0<\eps<<1$.
Then
\beq
\det( 1 + \eps A)
= 1 + \eps \tr(A) + O(\eps^2)
\;.
\eeq
\end{claim}
\proof
Just diagonalize $A$.
\qed

\end{document}